\documentclass[lettersize,journal]{IEEEtran}
\usepackage{amsmath,amsfonts}
\usepackage{algorithmic}
\usepackage{algorithm}
\usepackage{array}
\usepackage{multirow}
\usepackage{amsmath}
\usepackage[caption=false,font=normalsize,labelfont=sf,textfont=sf]{subfig}
\usepackage{textcomp}
\usepackage{stfloats}
\usepackage{booktabs}
\usepackage{flushend}
\usepackage{url}
\usepackage[hidelinks]{hyperref} 
\usepackage{verbatim}
\usepackage[table]{xcolor}
\usepackage{graphicx}
\usepackage{tabularx} 
\usepackage{color}
\usepackage{cite}
\begin{document}

\title{IGAA: Intent-Driven General Agentic AI for Edge Services Scheduling using Generative Meta Learning}

\author{Yan Sun, Yinqiu Liu, Shaoyong Guo*, Ruichen Zhang, Feng Qi, Xuesong Qiu,~\IEEEmembership{Senior Member,~IEEE,}\\ Weifeng Gong, Dusit Niyato,~\IEEEmembership{Fellow,~IEEE,} and Qihui Wu,~\IEEEmembership{Fellow,~IEEE}

\thanks{This work was supported by the National Natural Science Foundation of China (62322103). (*Corresponding author: Shaoyong Guo)}
\thanks{Yan Sun, Shaoyong Guo, Feng Qi, and Xuesong Qiu are with the State Key Laboratory of Networking and Switching Technology, Beijing University of Posts and Telecommunications, Beijing, 100876, China (E-mail: \{sunyan79, syguo, qifeng, xsqiu\}@bupt.edu.cn).}
\thanks{Yinqiu Liu, Ruichen Zhang, and Dusit Niyato are with the College of Computing and Data Science, Nanyang Technological University, 639798, Singapore (E-mail: yinqiu001@e.ntu.edu.sg, ruichen.zhang@ntu.edu.sg, DNIYATO@ntu.edu.sg).}
\thanks{Weifeng Gong is with the Inspur Computing Technology Pty Ltd, Beijing, 100095, China (E-mail: gongwf@inspur.com).}
\thanks{Qihui Wu is with the College of Electronic and Information Engineering, Nanjing University of Aeronautics and Astronautics, Nanjing 211106, China (E-mail: wuqihui@nuaa.edu.cn).}

}

\markboth{Journal of \LaTeX\ Class Files,~Vol.~14, No.~8, August~2021}%
{Shell \MakeLowercase{\textit{et al.}}: A Sample Article Using IEEEtran.cls for IEEE Journals}

\maketitle

\begin{abstract}
Agentic AI (AAI), which extends Large Language Models with enhanced reasoning capabilities, has emerged as a promising paradigm for autonomous edge service scheduling. However, user mobility creates highly dynamic service demands in edge networks, and existing service scheduling agents often lack generalization capabilities for new scenarios. Therefore, this paper proposes a novel Intent-Driven General Agentic AI (IGAA) framework. Leveraging a meta-learning paradigm, IGAA enables AAI to continuously learn from prior service scheduling experiences to achieve generalized scheduling capabilities. Particularly, IGAA incorporates three core mechanisms. First, we design a Network–Service–Intent matrix mapping method to allow agents to simulate novel scenarios and generate training datasets. Second, we present an easy-to-hard generalization learning scheme with two customized algorithms, namely Resource Causal Effect-aware Transfer Learning (RCETL) and Action Potential Optimality-aware Transfer Learning (APOTL). These algorithms help IGAA adapt to new scenarios. Furthermore, to prevent catastrophic forgetting during continual IGAA learning, we propose a Generative Intent Replay (GIR) mechanism that synthesizes historical service data to consolidate prior capabilities. Finally, to mitigate the effect of LLM hallucinations on scenario simulation, we incorporate a scenario evaluation and correction model to guide agents in generating rational scenarios and datasets.
Extensive experiments demonstrate IGAA's strong generalization and scalability. Specifically, IGAA enables rapid adaptation by transferring learned policies to analogous new ones, such as applying latency-sensitive patterns from real-time computing to optimize novel Internet of Vehicles (IoV) services. Compared to scenario-specific methods, IGAA maintains the intent-satisfaction rate gap within 3.81\%. In unseen scenarios, IGAA outperforms the best existing method by 19.19\%.

\end{abstract}


\section{Introduction}
\IEEEPARstart{A}{gentic} AI (AAI) represents a transformative paradigm for edge service scheduling, evolving traditional fixed schedulers into autonomous agents equipped with perception, reasoning, and adaptive decision-making capabilities \cite{ref33}. By integrating Large Language Models (LLMs) with various sensing and tool-execution modules, AAI can interpret service requirements, decompose complex scheduling objectives, and iteratively refine decisions through feedback. These characteristics make AAI a compelling solution for real-time resource management and dynamic service orchestration in mobile-edge computing \cite{ref1,ref2}.

Despite this potential, existing AAI proposals remain highly specialized and tailored to specific application scenarios\cite{ref53,ref54}. For example, an AAI designed for Internet of Vehicles (IoV) scenarios typically prioritizes mobility-aware decisions such as roadside unit selection, link stability maintenance, and predictive handover based on vehicular trajectories\cite{ref55}. Such agents encode decision logic optimized for high-mobility environments. However, when a Virtual Reality (VR) request enters the network, the IoV-oriented agent often fails to generate appropriate decisions, as VR scheduling depends on entirely different factors, including GPU capabilities and user immersion \cite{ref18}. This discrepancy underscores a fundamental limitation: \emph{current AAI cannot generalize across heterogeneous services and resource modalities}, especially in mobile-edge networks characterized by high user mobility, diverse service semantics, and rapidly evolving hardware capabilities \cite{ref17,ref19}.

To enhance the generality of agents, recent research has explored several technical directions. First, multi-agent and mixture-of-experts architectures decompose complex tasks into scenario-specialized agents and learn coordination policies among them, thereby broadening the range of solvable tasks \cite{ref34,ref35}. Although this improves coverage across scenarios, it also introduces non-negligible communication and coordination overhead, which is undesirable for latency-sensitive edge services.  
Second, modular and compositional policy learning improves generality by decomposing complex tasks into reusable sub-skills and recombining them to handle unseen scenarios\cite{ref36}. These methods allow agents to reuse previously learned behaviors, but they rely on a fixed skill library and cannot accommodate structural changes in the scheduling space, such as the introduction of new resource modalities or entirely new service types in edge networks.
Third, world model–based decision-making learns a shared latent representation of environment dynamics and uses it for scheduling in unseen tasks, enabling a certain degree of zero-shot or few-shot generalization\cite{ref38,ref39}. However, training a single world model that faithfully captures the coupled dynamics of diverse edge resources and services remains challenging in practice. Taken together, these directions are viewed as important building blocks toward Edge General Intelligence (EGI), but they do not yet provide a practical framework for general edge service scheduling.

In this paper, we exploit generative meta learning \cite{ref20} in AAI for edge service scheduling. 
Rather than training an agent on a single scheduling task, generative meta learning exposes the agent to a \emph{distribution} of heterogeneous tasks so that it can acquire task-invariant structures, such as shared resource dependencies and reusable scheduling behaviors across scenarios. 
By internalizing these cross-task regularities, a meta-learning–empowered agent can rapidly adapt to previously unseen service types and resource configurations with only minor parameter updates and training. 
Such fast adaptation is essential for mobile-edge networks, where both user tasks (e.g., IoV sensing, computation offloading and VR immersion) and available resources (e.g., CPU, GPU and Data Processing Unit (DPU) \cite{ref14}) evolve continuously and unpredictably.
However, implementing a meta-learning paradigm for AAI-based general service scheduling introduces several unique challenges.

\begin{itemize}
    \item \textbf{Challenge I: Dataset Construction}.  
    Generative meta learning requires a diverse set of training samples spanning multiple service scenarios. Dataset construction requires precisely converting high-level user intents into numerically valid resource demands \cite{ref21}. Although agents' LLMs excel at semantic understanding, they lack precise numerical reasoning aligned with physical environments such as GPU rendering latency, server workload, or wireless propagation modeling. 
    Consequently, generated samples often produce physically infeasible scenarios or unrealistic intent–resource mappings, misaligning with real network conditions.

    \item \textbf{Challenge II: Training Difficulty}.  
    As edge networks continuously incorporate new resource modalities and service types, the dimensionality of states and actions expands over time. 
    This introduces structural mismatches for conventional meta learning methods \cite{ref40} that assume fixed-dimensional inputs. 
    Additionally, training a single scheduling agent across heterogeneous scenarios often results in \emph{negative transfer} \cite{ref41}. Knowledge learned in one scenario can mislead learning in another scenario. Consequently, agents only capture scenario-specific behaviors instead of general scheduling principles. 
    Moreover, as the number of learned scenarios increases, the agent becomes increasingly vulnerable to \emph{catastrophic forgetting}, where adapting to new scenarios distorts or overwrites previously learned scheduling policies\cite{ref17}. 

\item \textbf{Challenge III: LLM Hallucinations}.  
To implement meta learning for AAI, LLMs play a central role, including understanding user intents, generating training samples, etc. 
However, LLM generation relies on stochastic token sampling driven by statistical correlations \cite{ref22}, which provides no guarantee of numerical consistency, logical coherence, or physical validity. 
Without explicit validation and correction, such hallucinations can propagate through the meta-learning pipeline, from dataset construction to cross-task training, ultimately leading to misleading supervision signals and suboptimal or invalid scheduling policies.

\end{itemize}

To overcome these challenges, we propose an Intent-Driven General Agentic AI (IGAA) framework that enables AAI to autonomously schedule general edge services. The key idea is to shift from task-specific policy training to an intent-centric meta-learning paradigm, where the agent continually learns new scheduling tasks generated from user intents while preserving and reusing knowledge learned from previous tasks. Our main contributions can be summarized as follows.
\begin{itemize}
    \item  We propose a Network–Service–Intent (N–S–I) matrix that converts natural-language semantics into numerical intent vectors. During dataset construction, the N–S–I matrix provides a physically grounded template that constrains LLM-generated scenarios, ensuring that the synthesized samples, such as latency targets, CPU/GPU/DPU demands, and service-specific resource patterns, remain numerically feasible and aligned with real network conditions \textbf{(for Challenge I)}. 
    \item  We design a meta-learning scheme in which agents gradually expand the task distribution by generating increasingly complex scheduling scenarios and modular training code, forming an easy-to-hard curriculum. To support efficient adaptation under this expanding task space, we develop two complementary algorithms: 1) Resource Causal Effect-aware Transfer Learning (RCETL), which identifies the resource dimensions whose causal influence changes when new hardware modalities appear, and selectively updates only the associated model parameters; and  2) Action Potential Optimality-aware Transfer Learning (APOTL), which transfers action-level priors learned from existing services to accelerate learning for newly introduced services. RCETL and APOTL jointly enable agents to adapt to new scenarios while preserving previously learned task-invariant behaviors. To address catastrophic forgetting, we further integrate a Generative Intent Replay (GIR), which synthesizes representative samples from past tasks, thereby reinforcing previously acquired scheduling capabilities throughout the meta-learning process \textbf{(for Challenge II)}.
    \item To address the hallucination issue, we develop a scenario evaluation–correction mechanism that audits LLM-generated datasets before adaptation. Crucially, this mechanism employs novel metrics as specific indicators to detect hallucinations by quantifying the deviation between generated scenarios and physical feasibility. When such inconsistencies are identified, the model automatically corrects or regenerates the corresponding data \textbf{(for Challenge III)}.
\end{itemize}

The remainder of this paper is organized as follows. Section II reviews the related works. Section III gives an overview of IGAA. Section IV presents the IGAA meta learning framework. Section V demonstrates the scenario generation process evaluation and correction model. Section VI discusses the experimental results and analysis. Finally, Section VII concludes this paper.

\section{Related Work}
\subsection{Agentic AI}
Agentic AI marks a paradigm shift from traditional task-oriented AI to autonomous, goal-driven AI agents with perception, multi-step reasoning, planning, and tool execution capabilities \cite{ref25, ref26}. In each AI agent, the LLM typically acts as the central controller, responsible for understanding user intents, decomposing tasks, and generating execution strategies. AAI vastly expands the boundaries of AI applications and achieves great success in various fields \cite{ref27,ref56}.

In the service scheduling scenario, AAI has already shown preliminary potential. Researchers have attempted to utilize AAI to perform automated fault diagnosis, network resource provisioning, and dynamic orchestration of service function chains. These works leverage in-context learning and Chain-of-Thought (CoT) of agents' LLMs to decompose complex tasks into executable API calls or scripts. For example, the authors in \cite{ref1} proposed a network management framework utilizing a network-specialized LLM. This system autonomously generates customized services through the collaboration of multiple scenario-specific agents. The authors in \cite{ref3} proposed a role-adaptive AAI-driven automated aerial vehicle swarm management method. In this method, the authors developed an LLM-driven semantic decision-making framework, using structured natural language for efficient semantic communication and collaborative reasoning. They also introduced a dynamic role heterogeneity mechanism for adaptive role switching and personalized management decisions. The authors in \cite{ref4} proposed a multi-agent collaborative AAI for automated large-scale network management. In this framework, a supervisor agent interprets user intents and coordinates specialized agents. The results are then validated through integration with a network simulation platform.

However, existing AAI frameworks for service scheduling are usually designed for specific scenarios \cite{ref17}. Especially in edge networks, the resource constraints of edge devices limit the parameter sizes of agents' LLMs, further affecting the generalizability. 
Moreover, previous work \cite{ref35, ref36, ref37, ref38, ref39, ref40} cannot fit the resource constraints, heterogeneity, and dynamics of mobile-edge environments.
Motivated by this, we propose a meta-learning framework for AAI to realize general service scheduling. 

\subsection{Intent-Driven Networking}
Intent-Driven Networking (IDN) is a network management paradigm. The core idea of IDN is to introduce a high-level abstraction layer that decouples users' Quality-of-Service (QoS) requirements from the corresponding network configurations \cite{ref23,ref61}. 
Intent is usually defined as a set of declarative, abstract, and vendor-agnostic rules that guide the service to satisfy the user's diverse needs. 
For example, a user can, while sending a service request, propose ``\textit{complete this computation task within 10 seconds}". 
IDN techniques can automatically translate this intent into the required amount of resources and perform configuration. 

With the application of LLM in network management, the IDN paradigm has received great attention from researchers. LLM, relying on powerful natural language understanding capabilities, can accurately translate the intent expressed by users through natural language into network configuration rules\cite{ref57,ref58}. For example, the authors in \cite{ref5} proposed an intent-driven service and resource management framework implemented based on LLMs. In this framework, the specialized agents collaborate to decompose the user intent and provide the computing infrastructure accordingly. The authors in \cite{ref6} utilize LLMs to convert natural language intent into network service descriptors and adopt a human feedback loop to learn from past experiences. The authors in \cite{ref7} use quantized low-rank adaptation to fine-tune LLMs, thus achieving memory-efficient intent processing. They also introduced a goal-aware hierarchical decision transformer to optimize service orchestration in the wireless access network.

To address the difficulty of converting natural-language intents into physically valid resource configurations, we design an N–S–I matrix as an intermediate semantic–numerical bridge.
The N–S–I matrix explicitly models the quantitative relationships among network resources, service types, and intent features, enabling Agentic AI to translate user intents into numerically grounded, executable resource requirements. This design serves two purposes.
First, it constrains the LLM-generated scenarios within realistic physical bounds (e.g., latency, bandwidth, or computing limits), ensuring the synthesized datasets remain consistent with real network behavior.
Second, it offloads fine-grained numerical reasoning from the LLM, allowing the model to focus on scenario generation and intent understanding rather than low-level computation.

\subsection{Edge General Intelligence}
EGI intends to deploy artificial general intelligence in heterogeneous edge computing environments \cite{ref8,ref59}. 
Traditional edge AI primarily focuses on compressing pre-trained, task-specific models for deployment on edge devices. In contrast, EGI aims to enable edge models to adapt to various service types, such as IoV and VR, as well as evolving resource constraints \cite{ref28}. 
For example, the authors in \cite{ref8} proposed a general intelligent wireless communication framework, which owns perception modules, world models, and action planning components to enhance generalization and handle unforeseen situations. 
Similarly, the authors in \cite{ref9} introduced an edge-adaptive AAI, enabling edge systems to autonomously reason and actively adapt through continuous perception-reasoning-action loops.

However, existing EGI proposals face two main challenges. First, given the diversity of resources and service types in edge networks, directly training agents across multiple scenario leads to negative transfer \cite{ref29}. Moreover, long-term learning suffers from catastrophic forgetting \cite{ref24}. To address the first challenge, we propose RCETL and APOTL, which improve learning efficiency by analyzing the correlations between previously learned scenarios and new ones. For the second challenge, we propose the GIR mechanism. In GIR, the LLM keeps sampling the previously learned scenario information, allowing IGAA to retain its learned service scheduling capabilities while learning new ones.

\section{Design Overview and Dataset Construction}
In this section, we introduce the basic workflow of IGAA for service scheduling (i.e., the inference phase), as well as the construction and updating of the N-S-I matrix.
\subsection{Overview of IGAA}
\begin{figure}[!t]
\centering
\includegraphics[width=3.5in]{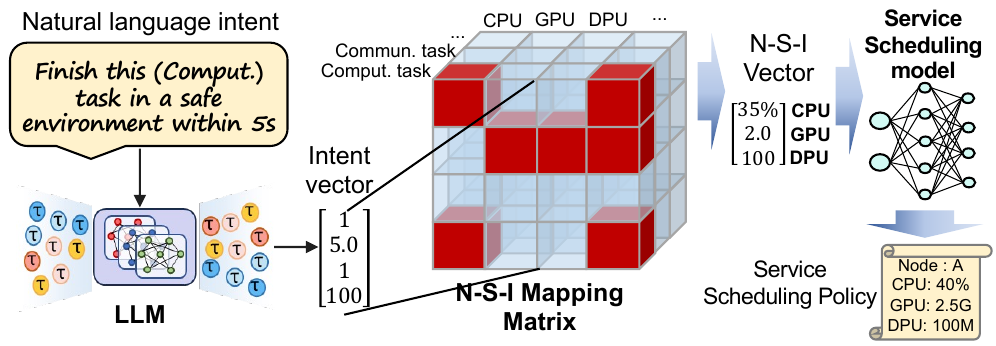}
\caption{Overview of IGAA (inference phase). IGAA transforms the natural language intent proposed by users into intent vectors and maps them into resource requirements vectors. Then, the service scheduling model deployed in IGAA generates personalized service scheduling policies accordingly.}
\end{figure}
In this paper, we consider a dynamic mobile edge network environment where users exhibit high mobility\cite{ref42}. The network supports a diverse combination of services, such as IoV, VR, and various IoT applications. In this context, the proposed IGAA framework functions as a general edge service scheduler that manages services within a given coverage area. It is capable of understanding users' high-level natural language intents and orchestrating heterogeneous edge services accordingly to meet specific QoS requirements\cite{ref6}.
IGAA consists of the following core modules:
\begin{itemize}
    \item \textbf{LLM}: IGAA uses an LLM (e.g., GPT or Deepseek) as the cognitive core. The LLM interprets user intents expressed in natural language and converts them into intent vectors\cite{ref4}. Moreover, during the transfer learning process, it continuously simulates new scenarios and generates the corresponding datasets.
    \item \textbf{N-S-I Matrix}: The N-S-I matrix transforms the intent vectors produced by the LLM into low-level resource requirements, thus facilitating the transfer learning. 
    \item \textbf{Service Scheduling Model}: This model generates service scheduling policies that satisfy user intents based on the derived resource requirements. Note that it is trained using RCETL and APOTL on the datasets generated by the LLM to acquire the generalization ability.
    \item \textbf{Evaluation and Correction Model}: After the service scheduling model has been trained, this module evaluates the rationale of the LLM-simulated scenarios using multiple metrics collected throughout the training process. It then guides the LLM to refine its scenario simulation and dataset generation\cite{ref22}.
\end{itemize}

The design of IGAA consists of two phases, namely inference and training. 
In this part, we primarily focus on the design of the inference phase. As shown in Fig. 1, during the inference phase, IGAA generates a service scheduling policy through the following steps:

\subsubsection{Intent Understanding} When the user issues a service request, they also provide a natural language expression of their intent, specifying the desired QoS requirements. The LLM analyzes the natural language intent and transforms it into an intent vector $I$. $I$ contains information such as the service type and QoS requirements. For example, in the intent vector shown in Fig. 1, the first element indicates a computing task, the second represents a latency requirement, and the third denotes the need for a secure environment. It is important to note that the intent vector is predefined, with each element having a specific meaning.

\subsubsection{Intent Vector Mapping} Based on the service type, IGAA selects an appropriate N-S-I matrix to map the intent vector $I$ into an N-S-I vector $N$. The construction of the N-S-I matrix is detailed in Section III. B. $N$ encodes the quantity of various network resources required to accomplish the task while satisfying the user’s intent. This process can be expressed as
\begin{equation}
    N=\boldsymbol{\Pi}_i^{(t)} \cdot I,
\end{equation}
where $i$ represents service type, and $\boldsymbol{\Pi}_i^{(t)}$ represents the N-S-I mapping matrix corresponding to type $i$ service at time $t$. As illustrated in the example in Fig. 1, the intent vector is transformed into an N-S-I vector, which specifies the required CPU, GPU, and DPU resources to complete the task within 5 seconds in a secure environment.
\subsubsection{Service Scheduling Policy Generation} Finally, the service scheduling model in IGAA generates a scheduling policy based on the N-S-I vector and the current network state, aiming to satisfy the user's intent. This policy includes decisions such as node selection and resource allocation.

\subsection{N-S-I Matrix Design and Dataset Generation}
Generating diverse service scheduling datasets poses a significant challenge to AAI due to the complex numerical reasoning required to map service types to specific requirements. To address this, we introduce an N–S–I matrix mechanism that bridges high-level intents with underlying network configurations, allowing the LLM in AAI to concentrate on intent understanding.
Specifically, the N-S-I Matrix transforms high-level user intents into quantifiable resource demands, providing directly usable inputs for the service scheduling model. As shown in Fig. 1, we define $\boldsymbol{\Pi}^{(t)} \in \mathbb{R}^{m\times n\times p}$ as the N-S-I Matrix at time $t$, where $m$ is the number of resource types, $n$ is the number of service types, and $p$ is the dimension of the intent vector. For a service of type $i$, its N-S-I submatrix $\boldsymbol{\Pi}_i^{(t)} \in \mathbb{R}^{m\times p}$ represents, in each column, the contribution weight of each element in the intent vector to the requirement of a specific resource. The coefficients of the initialized N-S-I Matrix are determined by the LLM through a combination of expert knowledge and historical execution data.

New scenarios often imply the introduction of new resources or new service types, which require the N-S-I matrix to be dynamically expanded.

\begin{figure*}[!t]
\centering
\includegraphics[width=\textwidth]{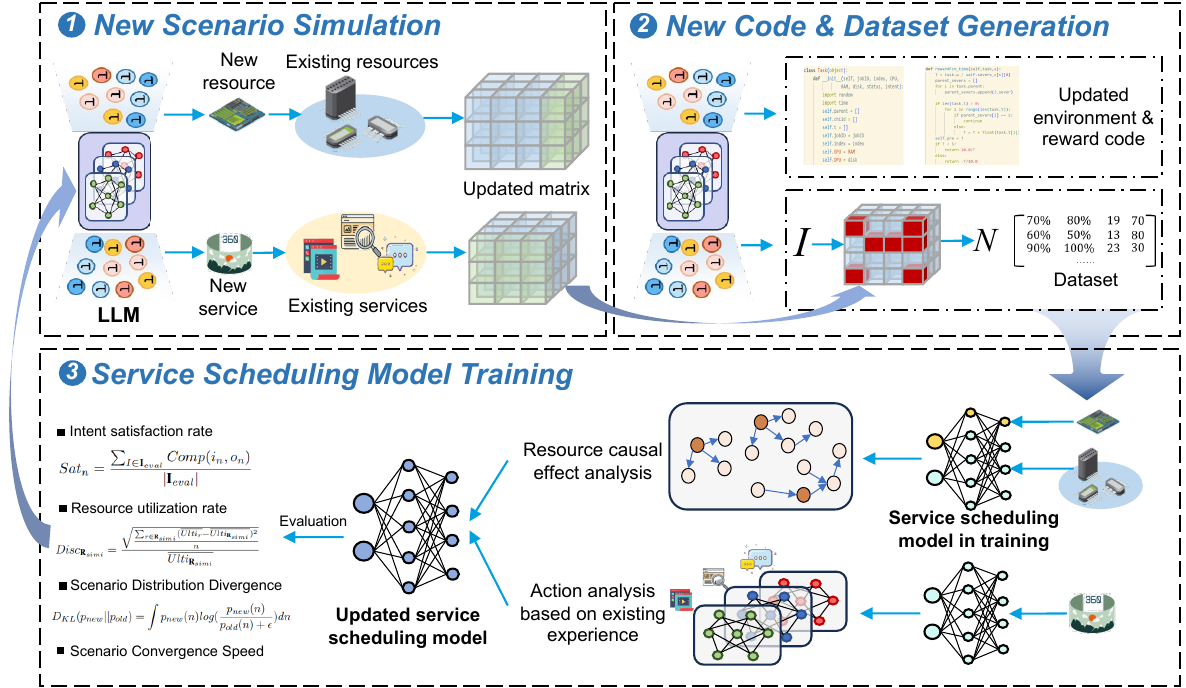}
\caption{Overview of IGAA (training phase). IGAA simulates new service scenarios according to CoT and updates the N-S-I matrix using the proposed algorithm. Then, IGAA generates new training code and datasets to enable the service scheduling model to acquire scheduling capabilities in the new scenario. Finally, the scheduling model performs transfer learning using the generated code and datasets. Besides, the scheduling model continuously interacts with the LLM to optimize the scheduling performance.}
\end{figure*}
\subsubsection{Introduction of New Resources}
When a set of $m'$ new resources is introduced (e.g., specialized hardware accelerators like DPUs added to existing CPUs), the matrix must be expanded along its first dimension. Let the current matrix be $\boldsymbol{\Pi}^{(t)} \in \mathbb{R}^{m \times n \times p}$. The LLM generates a new tensor $\boldsymbol{\Pi}_{\text{new\_res}} \in \mathbb{R}^{m' \times n \times p}$ representing the initial weights for these new resources. Concurrently, the weights for the original $m$ resources may be adjusted, represented by an adjustment function $g(\cdot)$. The updated matrix $\boldsymbol{\Pi}^{(t+1)} \in \mathbb{R}^{(m+m') \times n \times p}$ is formed by concatenating the adjusted original tensor and the new resource tensor along the $m$-dimension:
\begin{equation}
    \boldsymbol{\Pi}^{(t+1)} = \text{Concat}_{m}\left( g\left(\boldsymbol{\Pi}^{(t)}\right), \boldsymbol{\Pi}_{\text{new\_res}} \right),
\end{equation}
where $\text{Concat}_{m}(\cdot)$ denotes concatenation along the first (resource) dimension.

\subsubsection{Introduction of New Service Types} When $n'$ new service types are introduced, the matrix is extended along its second dimension. The LLM generates a new tensor $\boldsymbol{\Pi}_{\text{new\_svc}} \in \mathbb{R}^{m \times n' \times p}$. This initialization is typically guided by referencing the sub-matrices of existing similar services. We define this generation process as a function $f(\cdot)$ as follows
\begin{equation}
    \boldsymbol{\Pi}_{\text{new\_svc}} = f\left(\boldsymbol{\Pi}^{(t)}, S_{\text{similar}}\right),
\end{equation}
where $S_{\text{similar}}$ denotes the set of similar existing services. The updated matrix $\boldsymbol{\Pi}^{(t+1)} \in \mathbb{R}^{m \times (n+n') \times p}$ is then formed by concatenation along the $n$-dimension as follows
\begin{equation}
    \boldsymbol{\Pi}^{(t+1)} = \text{Concat}_{n}\left( \boldsymbol{\Pi}^{(t)}, \boldsymbol{\Pi}_{\text{new\_svc}} \right),
\end{equation}
where $\text{Concat}_{n}(\cdot)$ denotes concatenation along the second (service type) dimension.

During the IGAA training phase, IGAA first simulates the introduction of new network environments or new service types based on known service scenarios, as shown in Fig. 2. In this paper, we adopt the CoT paradigm to simulate new scenarios, where the complex simulation process of introducing new resources or new services is decomposed into a sequence of logically coherent intermediate reasoning steps. This structured reasoning guides the LLM to accurately update the N-S-I matrix and generate effective training datasets. We design two types of CoT to assist IGAA in scenario simulation and dataset generation, i.e.,
\begin{itemize}
    \item \textbf{CoT for New Network Environment Simulation:} Consider the integration of new resource types $\rightarrow$ assess the requirement for this resource across existing services $\rightarrow$ evaluate its impact on the current N-S-I mapping $\rightarrow$ update the mapping matrix accordingly.
    \item \textbf{CoT for New Service Type Simulation:} Consider the addition of new service types $\rightarrow$ identify the service intent associated with them $\rightarrow$ determine the types and amounts of resources required to fulfill this intent $\rightarrow$ update the mapping matrix accordingly.
\end{itemize}

Based on CoT and the procedures outlined in Section III. B, IGAA updates the N-S-I matrix. Then, it generates datasets to allow the service scheduling model to acquire scheduling capabilities in the new scenario. To do so, IGAA simulates user service requests and QoS intents in the new scenario, which are transformed through the N-S-I matrix.

Finally, the original service scheduling model performs transfer learning using the generated dataset to acquire new service scheduling capabilities. During training, the service scheduling model continuously interacts with the LLM. The LLM provides scenario correction and GIR to enhance the agent's performance (in Section IV.B). In return, the service scheduling model evaluates the generated scenario based on the evaluation model, thus avoiding generating unreasonable scenarios (in Section V).

\section{IGAA Meta Learning Framework}
As detailed in Section III, the LLM simulates novel scenarios and generates the corresponding training code and datasets. Building on these generated assets, this section presents the core meta-learning phase of the IGAA framework. With the dataset being constructed, in this section, we present two transfer learning algorithms in IGAA, namely RCETL and APOTL. Moreover, we demonstrate the GIR mechanism to defend against catastrophic forgetting.

\begin{figure*}[!t]
\centering
\includegraphics[width=5in]{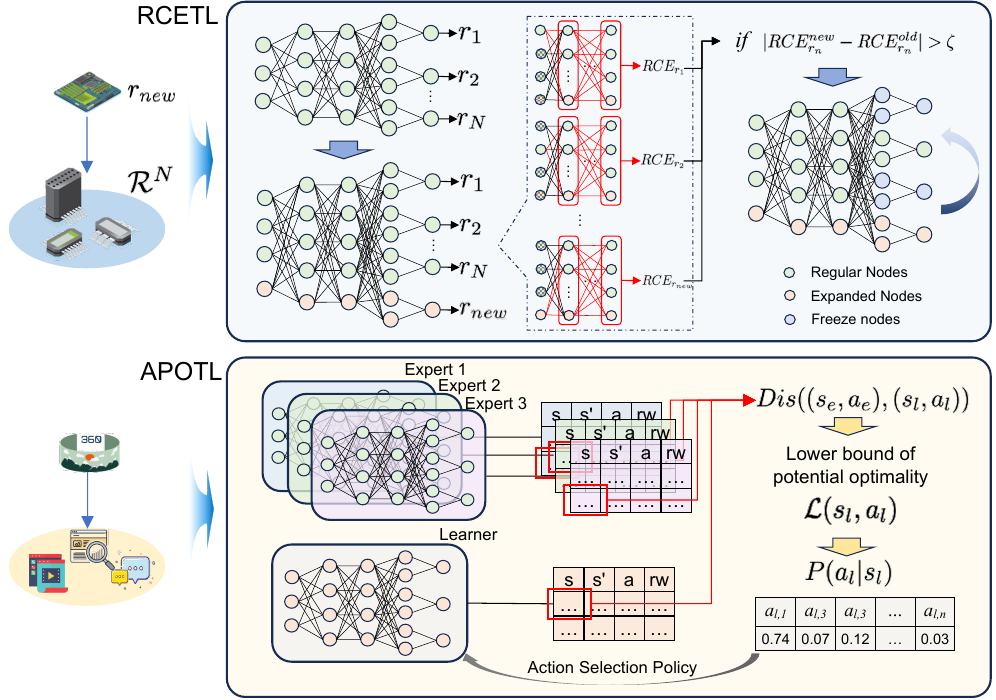}
\caption{RCETL aims to perceive the impact of newly introduced resources on existing ones and selectively fine-tune the parameters that are strongly correlated with the new resources. In APOTL, different sub-networks leverage prior service scheduling experience to optimize scheduling strategies when new service types are introduced. We derive the upper bounds of sub-network action potential optimality and use them to refine the service scheduling behavior selection strategy.}
\vspace{-3mm}
\end{figure*}

\begin{table}[tpb]
\renewcommand{\arraystretch}{1.5}
\caption{Main Notations\label{tab:table2}}
\centering
\begin{tabular}{p{2cm} p{5.5cm}}
\specialrule{2pt}{0pt}{0pt}
$\textbf{Notation}$ & $\textbf{Description}$\\
\hline
\hline
\multicolumn{2}{l}{Common Notations}\\
\hline
$E_{old},E_{new}$ & The encoder of old and new data in GIR mehcanism\\
$\textbf{I}$ & The set of test intent generated by LLM\\
$K_h$ & The Gaussian kernel function with bandwidth $h$\\
$KD(Dis)(A,B)$ & Kantorovich distance, quantifying the difference between distribution $A$ and $B$ under a specific distance metric $Dis$\\
$M_{\theta}$ & The service scheduling model with parameters $\theta$\\
$Net$ & The network state\\
$\mathbb{O},\mathbb{N}$ & The old and new data in GIR mechanism\\
$p_m,q_n$ & The state value of specific state $s$\\
$\textbf{R}_{simi}$ & The set of similar resources\\
$\mathcal{R}$ & The set of current network resource types\\
$r$ & The specific resource\\
$RCE$ & The resource causal effect vector\\
$Sat$ & The intent satisfaction rate\\
$s_i$ & The sample of resource requirements\\
$\Xi$ & The number of samples\\
\hline
\hline
\multicolumn{2}{p{\dimexpr 2cm+5.5cm+2\tabcolsep\relax}}{Parameters that need to be determined by the LLM in the equations}\\
\hline
$\eta_r$ ,$\eta_{KD}$ & The weights for the reward distance and the Kantorovich distance\\
$\xi$ & Causal strength threshold\\
\specialrule{2pt}{0pt}{0pt}
\end{tabular}
\end{table}

\subsection{IGAA Transfer Learning Algorithm Design}
Model adaptation exhibits different characteristics when introducing new resources or new services\cite{ref43}. Therefore, we design two specific transfer learning algorithms, namely RCETL and APOTL. As shown in Fig. 3, RCETL aims to perceive the impact of newly introduced resources on existing ones and selectively fine-tune the parameters that are strongly correlated with the new resources. In APOTL, different sub-networks leverage prior service scheduling experience to optimize scheduling strategies when new service types are introduced. We derive the upper bounds of sub-network action potential optimality and use them to refine the service scheduling behavior selection strategy. Table I lists the main notations. The system state transitions during IGAA training can be modeled as a Markov Decision Process (MDP) $(s, a, \tau)$, representing the system state, the action taken, and the reward value, respectively\cite{ref44}. We define the system state as $s = [Net, R_{req}]$, where $Net$ represents the network state, including the remaining resources in the system and the positions of the resource nodes, and $R_{req} \in \mathbb{R}^N$ represents the resource requirements of the service. The action is defined as $a = [ns, R_{alloc}]$, where $ns$ denotes the selected execution node for the service, and $R_{alloc}$ represents the amount of resources allocated to it.

\subsubsection{RCETL} 
Assume that the current set of network resource types is $\mathcal{R}^N$, where $N$ is the number of resource types. A pre-trained base model $M_\theta$ with parameters $\theta$ is available. Consider introducing a new network resource $r_{new}$ into the environment, that is, $\mathcal{R}^{N+1}=[\mathcal{R}^{N},r_{new}]$. We first define the Resource Causal Effect (RCE) value for each type of resource. For the resource requirement vector $R^N=[r_1,r_2,\ldots,r_N]$, we define the intervention RCE value of the feature $r_n$ as the difference between the expected intervention reward when $r_n$ is fixed to $a$, and the average expected reward of $r_n$:
\begin{equation}
    RCE_{r_n=a}=\left|\mathbb{E}\left(\tau|{\rm{Fix}}\left(r_n\!=\!a\right)\right)-\mathbb{E}_{r'}\left[\mathbb{E}\left[\tau|{\rm{Fix}}(r_n\!=\!a)\right]\right]\right|,
\end{equation}
where $\tau$ represents the reward of the service scheduling model. Fix$(\cdot)$ represents fixing a variable to a certain value. $r'\sim \mathcal{U}(l_n,h_n)$, where $l_n$ and $h_n$ denote the lower and upper bounds of $r_n$, respectively. The expected intervention reward is obtained by intervening on $r_n$ and computing the expected reward when other variables are uniformly sampled within their respective ranges:
\begin{equation}
\begin{aligned}
    & \mathbb{E}(\tau|{\rm{Fix}}(r_n=a)) = \int \tau \cdot p(\tau|{\rm{Fix}}(r_n=a))drw \\
    & = \int_{l_1}^{h_n} \dots \int_{l_{n-1}}^{h_{n-1}}\int_{l_{n+1}}^{h_{n+1}} \dots \int_{l_{N}}^{h_{N}} \Upsilon(r_1, \dots, r_N) \\
    & \quad \cdot p(\Upsilon|{\rm{Fix}}(r_n=a))dr_1 \ldots dr_{n-1}dr_{n+1} \dots dr_N,
\end{aligned}
\end{equation}
where $\Upsilon(\cdot)$ represents the reward function. The average expected reward of $r_n$ refers to the mean reward value when $r_n$ is uniformly sampled within its range. It is used to eliminate the influence of factors unrelated to causal effects on the RCE value, i.e.,
\begin{equation}
    \mathbb{E}_{r'}[\mathbb{E}[\tau|{\rm{Fix}}(r_n=a)]]=\int_{l_n}^{h_n}p(r')\mathbb{E}[\tau|{\rm{Fix}}(r_n=r')]dr'.
\end{equation}
For the $i$-th sample $s_i=[s_i^1,s_i^2,\ldots,s_i^N]$, we define the intervention RCE vector corresponding to $s_i$ as
\begin{equation}
    RCE_{s_i}=[RCE_{r_1=s_i^1},RCE_{r_2=s_i^2},\ldots,RCE_{r_N=s_i^N}].
\end{equation}
Furthermore, we can obtain the average RCE vector as 
\begin{equation}
    RCE=\frac{1}{\Xi}\sum_{i=0}^\Xi RCE_{s_i},
\end{equation}
where $\Xi$ denotes the number of samples. Thus, we obtain $RCE = [RCE_{r_1}, \ldots, RCE_{r_N}]$, where $RCE_{r_N}$ represents the average RCE value of the feature $r_n$, indicating the degree of influence of the feature $r_n$ on resource scheduling.

For the original network resource set $\mathcal{R}^N$ and the updated network resource set $\mathcal{R}^{N+1}$, their corresponding RCE vectors are calculated as $RCE_{r_n}^{\rm{old}}$ and $RCE_{r_n}^{\rm{new}}$, respectively. For a feature $r_n$, if
\begin{equation}
    \xi_{r_n}=|RCE_{r_n}^{\rm{old}}-RCE_{r_n}^{\rm{new}}|>\xi,
\end{equation}
$r_n$ and $r_{new}$ are determined to be strongly causally related resources\cite{ref45}. Where $\xi$ is a hyperparameter representing the causal strength threshold, which can be dynamically adjusted during model training. During training, modules related to weakly causal resources are frozen to improve training efficiency\cite{ref46}. RCETL is summarized in Algorithm 1.
\begin{algorithm}[tpb]
\caption{RCETL algorithm.}
\begin{algorithmic}
\STATE 
\STATE {\textbf{Require}: The set of network resource types $\mathcal{R}^N$, existing model $M_{\theta}$, new network resource type $r_{new}$, causal strength threshold $\xi$, resource requirements of services $\textbf{Req}=[R_1^{N+1},\ldots,R_\Xi^{N+1}]$}
\STATE \textbf{Procedure}:
\STATE for $r$ in $\mathcal{R}^N$
\STATE \hspace{0.5cm} Calculate $RCE_{r_n}^{old}$ based on Eq. (5-9)
\STATE end for
\STATE Expand $M_\theta$ to form $M_{\theta'}$ to adapt to the new input and initialize the new parameters
\STATE for $r$ in $[\mathcal{R}^N,r_{new}]$ 
\STATE \hspace{0.5cm} Calculate $RCE_{r_n}^{new}$ based on Eq. (5-9)
\STATE \hspace{0.5cm}if $|RCE_{r_n}^{old}-RCE_{r_n}^{new}|<\xi$
\STATE \hspace{1cm} Freeze the parameters related to $r$ in $M_{\theta'}$ 
\STATE end for
\STATE while not converged do
\STATE \hspace{0.5cm} Generate service scheduling result by $a=M_{\theta'}(s)$
\STATE \hspace{0.5cm} Achieve the next state $s'$
\STATE \hspace{0.5cm} Storage $(s,s',a,\tau)$ in the replay buffer
\STATE \hspace{0.5cm} Update $\theta'$
\STATE end while
\end{algorithmic}
\label{alg1}
\end{algorithm}

\subsubsection{APOTL} As shown in Fig. 3, IGAA employs the APOTL method to leverage the experience of existing expert sub-networks when a new service type is introduced. By evaluating the potential optimality of the new sub-network, it improves the learning efficiency of the service scheduling capability. We first observe the distribution difference between the new learner sub-network and the expert sub-network. Let $D_e$ and $D_l$ represent the distributions of the expert sub-network and the new sub-network, respectively, and $\textbf{S}_e$ and $\textbf{S}_l$ denote their corresponding state sets. We use the Kantorovich distance to quantify the difference between the two distributions \cite{ref10}:
\begin{equation}
\begin{aligned}
    &KD(Dis)(D_e,D_l)\max\limits_{p_m,q_n}\left(\sum_m^{|\textbf{S}_e|}D_e(s_m)p_m-\sum_n^{|\textbf{S}_l|}D_e(s_n)q_n\right), \\
    & \quad\quad\quad\quad\quad s.t. (C1)\ p_m-q_n \le Dis(s_m,s_n),\\
    &\quad\quad\quad\quad\quad\;\;\;\;\;\;(C2)\ -1 \le p_m \le1, \\
    &\quad\quad\quad\quad\quad\;\;\;\;\;\;(C3)\ s_m \in \textbf{S}_e,s_n \in \textbf{S}_l ,
\end{aligned}
\end{equation}
where $Dis$ represents a specific distance metric. $p_m$ and $q_n$ are internal optimization parameters, representing a specific state value when the state is $s$. The constraint (C1) limits the distance between these two values, i.e., it enforces the Lipschitz constraint \cite{ref10}. The Kantorovich distance shows the difference between the two distributions under the metric $Dis$. Intuitively, it represents the minimal cost required to transform distribution $D_l$ into $D_e$.

Assume that the expert sub-network and the new sub-network have MDPs represented by $<s_e,s_e',a_e,\tau_e>$ and $<s_l,s_l',a_l,\tau_l>$, respectively. We use the lax bisimulation metric \cite{ref10} to represent the distance between their state-action pairs:
\begin{equation}
\begin{aligned}
    &Dis\left((s_e,a_e),(s_l,a_l)\right)=\eta_r|\tau_e(s_e,a_e)-\tau_l(s_l,a_l)|\\
    &+\eta_{KD}(KD(Dis'(D_e(s_e,a_e),D_l(s_l,a_l)))),   
\end{aligned}
\end{equation}
where $\eta_r$ and $\eta_{KD}$ represent the weights for the reward distance and the Kantorovich distance, respectively, which can be dynamically adjusted by the LLM during training. In this work, we use the Hausdorff metric as the distance measure in the Kantorovich distance. It represents the maximum distance from a point in one set to the nearest point in another set. This creates a nested metric structure where the Hausdorff distance quantifies the geometric deformation cost between individual samples, and the Kantorovich distance aggregates these costs to evaluate the global discrepancy between their probability distributions\cite{ref10}. By considering both the worst-case scenario and the nearest neighbor, it ensures that the action sets remain similar even in the worst case:
\begin{equation}
\begin{aligned}
    &Dis'(D_e(s_e,a_e),D_l(s_l,a_l))\\
    &=\max\left(\max\limits_{a_e\in \textbf{A}_e}\min\limits_{a_l\in \textbf{A}_l}Dis(s_e,a_e),(s_l,a_l),\right.\\
    &\quad \quad \quad \;\;\left.\min\limits_{a_e\in \textbf{A}_e}\max\limits_{a_l\in \textbf{A}_l}Dis(s_e,a_e),(s_l,a_l)\right).
\end{aligned}
\end{equation}

We then use the above lax bisimulation to evaluate the potential optimality of behaviors. APOTL is trained using the DDQN algorithm, aiming to maximize the long-term expected reward, i.e., the optimal state value under state $s$\cite{ref30}:
\begin{equation}
    V(s)=\mathbb{E}\left(\sum_{i=0}^{\infty}{\gamma^i r(s_i,a_i)|s_0}\right),
\end{equation}
where $\gamma$ is the discount factor. Meanwhile, the model updates the Q-value, which represents the expected long-term cumulative reward obtained after taking a specific action:
\begin{equation}
    Q(s,a)=r(s,a)+\gamma\sum_{s\in\textbf{S}}{Q(s'|s,a)V(s')}),
\end{equation}
where $s'$ represents the next state, and $\alpha$ is the learning rate. To evaluate the potential optimality of the learner sub-network leveraging the expert’s behavior, we first calculate the state value difference \cite{ref10}:
\begin{equation}
\begin{aligned}
    &\left|Q_l^*(s_l,a_l)-V_e^*\right|\\
    &=\left|Q_l^*(s_l,a_l)-Q_e^*(s_e,a_e^*)\right|\\
    &=\left| r_l(s_l,a_l)+\gamma\sum_{s_l\in\textbf{S}_l}Q_l(s_l'|s_l,a_l)V(s_l') \right.\\
    &\quad \left. -r_e(s_e,a_e^*)+\gamma\sum_{s_e\in\textbf{S}_e}Q_e(s_e'|s_e,a_e)V(s_e') \right|\\
    &\le \left|r_l(s_l,a_l)-r_e(s_e,a_e^*)\right|\\
    &\quad +\left|\gamma\sum_{s_l\in\textbf{S}_l}Q_l(s_l'|s_l,a_l)V(s_l')-\gamma\sum_{s_e\in\textbf{S}_e}Q_e(s_e'|s_e,a_e^*)V(s_e')\right|\\
    &\le \left|r_l(s_l,a_l)-r_e(s_e,a_e^*)\right|+\max\limits_{a_l\in \textbf{A}_l}\min\limits_{a_e\in \textbf{A}_e}\left( \right.\\
    &\quad \left. \gamma \left|\sum_{s_l\in\textbf{S}_l}Q_l(s_l'|s_l,a_l)V(s_l')-\sum_{s_e\in\textbf{S}_e}Q_e(s_e'|s_e,a_e^*)V(s_e')\right| \right)\\
    &= \left|r_l(s_l,a_l)-r_e(s_e,a_e^*)\right|+\gamma KD(Dis')((s_e,a_e^*),(s_l,a_l))\\
    &\quad \quad (\text{using Eq. (11) and Eq. (13)})\\
    &= Dis((s_e,a_e^*),(s_l,a_l)), \quad (\text{using Eq. (12)})
\end{aligned}
\end{equation}
where $Q^*(s_l, a_l)$ represents the optimal Q-value of $(s_l, a_l)$ in the new sub-network. $V^*(s_e)$ denotes the optimal state value in the expert network, where $s_e$ is the state in the expert network closest to $s_l$, and $a_e^*$ is the optimal action under state $s_e$. $s_l’$ represents the next state after taking action $a_l$ in state $s_l$. The Eq. (16) indicates that the value difference between $(s_l, a_l)$ and $(s_e, a^*_e)$ has an upper bound, based on which we can derive the upper bounds of the behavioral potential optimality of the learner sub-network.

During training, APOTL typically uses $V^*(s_l)$ as the target for $Q^*(s_l, a_l)$. Therefore, we compute the potential optimality of the learner sub-network using the difference $V^*(s_l) - Q^*(s_l, a_l)$. We can derive the upper bound of the potential optimality when selecting action $a_l$ under state $s_l$:
\begin{equation}
\begin{aligned}
    &|V^*_l(s_l)-Q^*_l(s_l,a_l)|\\
    &=|Q^*(s_l,a_l^*)-Q_l^*(s_l,a_l)|\\
    &=|Q^*(s_l,a_l^*)-V_e^*(s_e)+V_e^*(s_e)-Q_l^*(s_l,a_l)|\\
    &\le|Q^*(s_l,a_l^*)-V_e^*(s_e)|+|V_e^*(s_e)-Q_l^*(s_l,a_l)|\\
    &\le Dis((s_l,a_l^*),(s_e,a_e^*))+Dis((s_l,a_l),(s_e,a_e^*))\\
    &=UP_e(s_l,a_l).
\end{aligned}
\end{equation}

This upper bound indicates that, by leveraging expert experience, the gap between the action selected by the subnetwork and the optimal action is limited. We aim to choose actions with smaller upper bounds, as they possess greater potential advantages. Since $a_l^*$ and $a_e^*$ are fixed, the upper bound depends only on $s_l$ and $a_l$. Considering that multiple expert networks may exist in the model, we denote $UP(s_l, a_l)$ as the upper bound of the potential optimality of action $a_l$, which can be calculated as follows:
\begin{equation}
   UP(s_l,a_l)=\mu_1{UP_{e_1}}(s_l,a_l)+\cdots+\mu_n{UP_{e_n}}(s_l,a_l),
\end{equation}
where $n$ is the total number of expert networks, and $\mu$ represents the influence weight of each expert network, satisfying $\mu_1+\mu_2+\ldots+\mu_n=1$. The LLM can determine these weights before training by assessing the similarity between the new service type and existing ones, and continuously optimize them during the training process.

Finally, we can determine the action selection strategy based on the potential optimality. Since we aim to choose actions with higher potential advantage, which means lower $UP(s_l,a_l)$, the action selection probability can be calculated as follows:
\begin{equation}
    P(a_l|s_l)=\frac{Sigmoid(-UP(s_l,a_l))}{\sum_{a\in \textbf{A}_l}Sigmoid(-UP(s_l,a))}.
\end{equation}

Through the above method, we can utilize the experience of expert networks to guide the action selection of the new sub-network, thereby improving the training efficiency of the sub-network. The proposed APOTL algorithm is summarized in Algorithm 2.
\begin{algorithm}[t]
\caption{APOTL algorithm.}
\begin{algorithmic}
\STATE 
\STATE {\textbf{Require}: The set of existing expert sub-network $[e_1,e_2,\ldots,e_n]$, sub-network of learner $\mathcal{L}$, the state set and action set of expert $[\textbf{S}_{e_1},\ldots,\textbf{S}_{e_n}]$ and $[\textbf{A}_{e_1},\ldots,\textbf{A}_{e_n}]$, resource requirement of services $\textbf{Req}=[R_1^{N},\ldots,R_\Xi^{N}]$}
\STATE \textbf{Procedure}:
\STATE Init parameters of $\mathcal{L}$
\STATE while not converged do
\STATE \hspace{0.5cm} Generate action set $\textbf{A}_l=\mathcal{L}(Net, R^N)$ 
\STATE \hspace{0.5cm} for $a_l$ in $\textbf{A}_l$ 
\STATE \hspace{1cm} Calculate $Dis((s_e,a_e^*),(s_l,a_l))$ using Eq. (11-13)
\STATE \hspace{1cm} Calculate upper bound of the potential optimality of\\ \quad\quad\,\;\;\, action $a_l$ using Eq. (17)
\STATE \hspace{0.5cm} end for
\STATE \hspace{0.5cm} Calculate action selection probability using Eq. (19)
\STATE \hspace{0.5cm} $a=argmax(P(a_l|s_l))$ 
\STATE \hspace{0.5cm} Storage $(s,s',a_l,\tau)$ in replay buffer 
\STATE \hspace{0.5cm} Update the parameters of $\mathcal{L}$ 
\STATE end while
\end{algorithmic}
\label{alg2}
\end{algorithm}

\subsubsection{Complexity Analysis}
The complexity of the RCETL algorithm mainly consists of RCE computation and subsequent model training. The RCE computation is used to determine the causal relationships between new and existing resources. This computation involves sampling $\Xi$ instances and evaluating $N$ resource types. Assuming the computational cost of evaluating the reward function is $C_{\Upsilon}$, the precomputation complexity of RCE is approximately $\mathcal{O}(\Xi \times (N+1) \times C_{\Upsilon})$ \cite{ref48}. Assume the service scheduling model $M_{\theta'}$ is a neural network with $L_{rcetl}$ layers and an average size of $h_{rcetl}$. Let the batch size be $B$ and the proportion of unfrozen parameters be $f$ ($0 < f \le 1$). Then the complexity per batch (one forward pass and one backward pass) is approximately $\mathcal{O}(B \times f \times L_{rcetl} \times h_{rcetl}^2)$ \cite{ref47}. Inference (i.e., generating one action) requires one full forward pass through $M_{\theta'}$, with complexity $\mathcal{O}(L_{rcetl} \times h_{rcetl}^2)$.

The complexity of the APOTL algorithm is mainly determined by its action selection process, which uses the experience of $n$ expert networks to evaluate the potential optimality of each action. Assume the learner network $L$ has $L_{apotl}$ layers and an average size of $h_{apotl}$. For a given state $s_l$, the size of the discretized action set is $|\textbf{A}_l|$. Let the complexity of computing a single Kantorovich distance be $C{KD}$. Therefore, the complexity of selecting an optimal action at a single time step (i.e., inference complexity) is $\mathcal{O}(|A_l| \times n \times C_{KD})$\cite{ref48}. The training complexity is the sum of the action selection and model parameter update costs. The network update complexity is approximately $\mathcal{O}(B \times L_{apotl} \times h_{apotl}^2)$. Hence, the total training complexity per batch is $\mathcal{O}(B \times ( (|A_l| \times n \times C_{KD}) + (L_{apotl} \times h_{apotl}^2) ))$ \cite{ref49}. Based on this analysis, the inference complexity is $\mathcal{O}(|A_l| \times n \times C_{KD})$.

\begin{figure}[!t]
\centering
\includegraphics[width=3.5in]{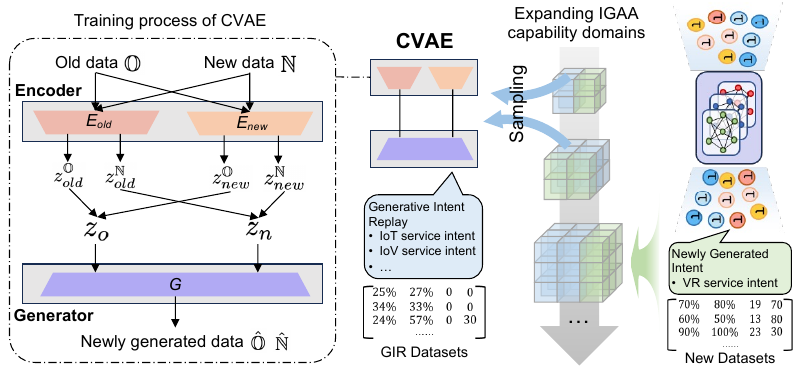}
\caption{Overview of Generative Intent Replay. IGAA samples earlier simulated service scenarios and generates the corresponding service intents. These intents are then combined with those from new scenarios to form new dataset.}
\end{figure}

\subsection{Generative Intent Replay}
Continuously learning new service scheduling strategies may cause early strategies to become invalid, leading to catastrophic forgetting, which degrades IGAA’s performance in certain scenarios. Even if early datasets are preserved and merged with new ones, this may result in overfitting to earlier tasks\cite{ref11}. To address this issue, we propose the GIR mechanism. As shown in Fig. 4, the GIR model samples earlier service request data and generates the corresponding service intents, which are then combined with the intents from new scenarios to form an augmented dataset. In our framework, we employ a Variational Autoencoder (VAE)\cite{ref31} as the GIR model. It should be noted that although the LLM itself could serve as a GIR model, doing so would introduce additional content generation processes, reducing IGAA's training efficiency. Moreover, frequent context switching might increase the likelihood of hallucinations. By introducing a separate GIR model that expands synchronously with IGAA, we ensure that the LLM remains focused on generating training code and datasets for IGAA\cite{ref50,ref51}. In addition, the computational cost of the VAE model is much lower than that of the LLM, so the inclusion of a dedicated GIR model does not cause excessive computational overhead.

The Contrastive VAE (CVAE)-driven GIR process we designed is shown in Fig. 4. In each round, when IGAA trains a new service scheduling capability, that is, when the LLM generates a new scenario, the CVAE simultaneously generates a GIR dataset. The GIR dataset imitates the data distribution of service demands in existing scenarios to simulate user intents in those scenarios. The GIR dataset and the new dataset generated by the LLM are mixed to form the training dataset for this round of IGAA training. This allows IGAA to learn new service scheduling capabilities while retaining its original ones. 

To enable CVAE to learn the data distribution of service demands in new scenarios, it is trained to generate new types of data concurrently with IGAA's training of new service scheduling capabilities. As shown in the figure, the CVAE consists of a new data encoder $E_{new}$, an old data encoder $E_{old}$, and a generator $G$. In each training round, we first freeze the new data encoder $E_{new}$ and train $E_{old}$ with old data, then freeze $E_{old}$ and train $E_{new}$ with new data. The loss functions for the two encoders are respectively defined as follows:
\begin{equation}
    L_{old}=MSE+KL(p(z_{old}^\mathbb{O})||p(z)),
\end{equation}
\begin{equation}
    L_{new}=MSE+KL(p(z_{new}^\mathbb{N})||p(z)),
\end{equation}
where $MSE$ denotes the mean squared error function, and $KL(\cdot)$ represents the Kullback–Leibler (KL) divergence between the feature distribution generated by the encoder and the target distribution. Then, we freeze $E_{old}$ and $E_{new}$, and train the generator with an equal proportion of new and old data. The loss function is defined as follows:
\begin{equation}
    L_{G}=MSE+KL(p(z_{o}||p(z))+KL(p(z_{n}||p(z)).
\end{equation}

It should be noted that during generator training, we do not simply mix the encoded new and old data. Instead, both new and old data are processed once by $E_{old}$ and $E_{new}$, respectively, and the resulting features are stacked to form mixed features $z_o$ and $z_n$. This approach enables the generator to more effectively analyze the similarities between new and old data and to integrate these similarities to generate higher-quality data\cite{ref12}.

\section{Scenario Generation Process Evaluation and Correction Model}
In this section, we propose an evaluation and correction model for the service scenario generation process. By integrating this model, the service scheduling model can evaluate the rationality of the LLM’s scenario generation, training code, and dataset creation through multiple metrics. These evaluations then guide the LLM to optimize its training process. The model primarily consists of indicators such as intent satisfaction rate, homogeneous resource utilization, scenario distribution divergence, and scenario convergence speed.

\subsection{Intent Satisfaction Rate}
The intent satisfaction rate directly measures whether the scenarios generated by the LLM are reasonable. The LLM might generate an unsatisfiable intention, such as requiring extremely low latency in a resource-constrained environment. Alternatively, it might produce a reward function that cannot properly guide the service scheduling model to learn new strategies. In such cases, no matter how the service scheduling model is trained, it cannot achieve a high intent satisfaction rate.

After the service scheduling model completes training for a new scenario, the LLM generates a set of natural language intents $\textbf{I}_{\text{eval}}$ for testing under that scenario. Each $I \in \textbf{I}_{\text{eval}}$ contains multiple fine-grained intents in different aspects, i.e., $I = [i_1, i_2, \ldots, i_n]$, simulating users' diverse QoS requirements. For each intent $i \in I$, the trained model generates a service scheduling policy, which is executed in a simulated environment to obtain the actual service execution outcomes $O_I = {o_1, o_2, \ldots, o_n}$. We define $Comp(i_n, o_n) = 1$ if the fine-grained intent $i_n$ is satisfied, and $Comp(i_n, o_n) = 0$ otherwise. After evaluating all test intents in $\textbf{I}_{\text{eval}}$, we calculate the satisfaction rate for each fine-grained intent:
\begin{equation}
{Sat}_n=\frac{\sum_{I\in \textbf{I}_{eval}}{Comp(i_n,o_n)}}{|\textbf{I}_{eval}|}.
\end{equation}

A low intent satisfaction rate (in this paper we set this threshold to 0.75 \cite{ref52}) guides the LLM to re-examine its scenario generation reasoning or the reward settings in the training code, in order to improve the model’s performance on specific intents.

\subsection{Utilization Rate of Similar Resources}
The utilization rate of similar resources is used to address the problem of over-reliance on a specific type of resource. During the training of some scenarios, the service scheduling model may focus solely on the reward value. Although relying on a single resource may be sufficient to complete the task and obtain a reward, in practice, this could exacerbate resource contention. A reasonable scenario should not only be solvable but also guide the service scheduling model to learn efficient and robust strategies rather than depending on a single type of resource. We define multiple resources that can achieve the same function and assist each other to achieve higher performance as similar resources. For example, both a CPU and a DPU can perform basic data processing. In a node equipped with both, the CPU can offload part of the task to the DPU to relieve computational pressure\cite{ref60}. Therefore, we consider the CPU and DPU as similar resources.

Assume that the set of nodes in the environment is $\textbf{M}$, with the number of nodes being $r_m$. Similar to the steps in subsection V. A, the LLM generates test intents. We assume that, after service scheduling is completed, the utilization of resource $r_n$ at each node is $Ulti_{m,r_n}$. For a set of similar resources $\textbf{R}_{simi} = [r_1, r_2, \ldots, r_n]$, the dispersion coefficient of similar resources can be calculated as follows:
\begin{equation}
Disc_{\textbf{R}_{simi}}=\frac{\sqrt{\frac{\sum_{r \in \textbf{R}_{simi}}(\overline{Ulti_r}-\overline{Ulti_{\textbf{R}_{simi}}})^2}{n}}}{\overline{Ulti_{\textbf{R}_{simi}}}},
\end{equation}
\begin{equation}
    \overline{Ulti_r}=\frac{1}{m}\sum_{m \in \textbf{M}}Ulti_{m,r},
\end{equation}
\begin{equation}
    \overline{Ulti_{\textbf{R}_{simi}}}=\frac{1}{n_r}\sum_{r \in \textbf{R}_{simi}}\overline{Ulti_{r}}.
\end{equation}

First, Eq. (26) computes the average utilization of a single type of resource across all network nodes. Then, the overall average utilization level of a group of similar resources is obtained through Eq. (25). Finally, Eq. (24) calculates the normalized standard deviation of each resource’s utilization relative to the group’s average level.

A high dispersion coefficient indicates whether the LLM can achieve the same effect using similar types of resources. Upon receiving this feedback. The LLM may adjust the weights of relevant resources in the N-S-I matrix, or deliberately create bottlenecks in over-relied resources when generating new scenarios. This strategy forces the service scheduling model to learn to use alternative resources.

\subsection{Scenario Distribution Divergence}
Scenario distribution divergence is used to evaluate the novelty and diversity of the scenario generated by the LLM. This helps prevent the LLM from falling into mode collapse, where it repeatedly generates similar scenarios and datasets.
In each round, when the LLM generates new scenarios and produces a new set of intent $\textbf{I}$, we model this as a data distribution $p_{\text{new}}$:
\begin{equation}
    p_{new}(n)=\frac{1}{|\textbf{I}|}\sum_{I\in \textbf{I}}{K_h(n-I)},
\end{equation}
where $K_h$ is a Gaussian kernel function with bandwidth $h$\cite{ref13}. Similarly, from the historical intent vector sets $\{\textbf{I}_1, \textbf{I}_2, \dots\}$, we extract the historical data distributions $[p_{\text{old}, \textbf{I}_1}, p_{\text{old}, \textbf{I}_2}, \dots]$.

Then, we use the KL divergence to assess the similarity between the scenarios:
\begin{equation}
    D_{KL}(p_{new}||p_{old})=\int p_{new}(n)log\left(\frac{p_{new}(n)}{p_{old}(n)+\epsilon}\right)dn,
\end{equation}
where $\epsilon$ is a very small value to prevent the denominator from being zero. We use Monte Carlo sampling, directly employing the samples in I to approximate this KL divergence:
\begin{equation}
    D_{KL}(p_{new}||p_{old})\approx\frac{1}{|\textbf{I}|}\sum_{I \in \textbf{I}}[log(p_{new}(I))-log(p_{old}(I)+\epsilon)].
\end{equation}

For any historical intent vector set, if $D_{KL}(p_{new}||p_{old})$ is below a certain threshold, the newly generated scenarios by the LLM are considered insufficiently novel, which prompts the LLM to regenerate new scenarios.

\subsection{Scenario Convergence Speed}
Considering that this paper proposes an easy-to-hard framework for learning edge service scheduling capabilities, we do not want the LLM to simulate an overly advanced scenario. Instead, we aim for it to incrementally add a small number of new resource or service types based on existing scenarios. This approach is designed to avoid low learning efficiency or forgetting historical capabilities in the service scheduling model.

If $Sat_n$ and $Disc_{\textbf{R}_{simi}}$ show no anomalies after training, we measure the steps required to reach the performance threshold. In this paper, we use $Sat_n > 90\%$ as the performance threshold. A very short convergence time typically indicates that the new scenario is relatively simple, similar to cases where $D_{KL}(p_{new}||p_{old})$ falls below a certain threshold. Consequently, this prompts the LLM to generate more novel scenarios. Conversely, an excessively long convergence time, or failure to converge, indicates that the service scheduling difficulty of the new scenario is too high and the model cannot effectively learn it. In such cases, the LLM should reduce the difficulty of the scenario or decompose it into several simpler intermediate scenarios\cite{ref32}.

\subsection{Closed-Loop Correction and Optimization}
Based on the multi-dimensional metrics defined above, IGAA implements a holistic reflection and correction loop to ensure the validity of the generated scenario and datasets. After training of service scheduling model in new scenario, the evaluation and correction model aggregates the above indicators and report them to LLM. This enables the LLM to reason about the root causes of training failures, whether they stem from physically infeasible scenarios (hallucinations), biased resource initialization in the N-S-I matrix, or ineffective reward guidance. Guided by this high-level feedback, the LLM autonomously regenerates or refines the scenario descriptions, N-S-I mapping coefficients, and training code, thereby iteratively aligning the simulated learning curriculum with the agent’s actual processing capabilities. For example, consider a case where the LLM generates a physically infeasible intent, such as demanding sub-millisecond latency for a heavy computing task in a bandwidth-constrained IoT environment. The evaluation model detects that the intent satisfaction rate remains critically low despite extensive training cycles, signaling a validity gap. This feedback serves as a correction trigger, prompting the LLM to revise the generated scenario by relaxing the latency constraints. Furthermore, we give some more numerical results to support our proposal.

\section{Numerical Results}
In this section, we implement the proposed IGAA framework, including its training and inference processes. We conduct extensive experiments to evaluate the generality of IGAA and verify the effect of the RCETL algorithm, APOTL algorithm, GIR component, and the evaluation-and-correction model proposed in the IGAA framework.

\textbf{Testbed}: IGAA training and testing were conducted on a server equipped with an AMD EPYC 7763 CPU, 4$\times$ GeForce RTX 4090 (24GB), running Ubuntu 18.04.

\textbf{Experimental Settings}: During IGAA training, the Claude-4.5-Sonnet model was used to generate training code and datasets for the service scheduling model. During testing, we deploy the DeepSeek-R1 (7B) model to understand user service intents. The service scheduling model was trained using PyTorch 2.5.0.

\textbf{Baselines}: In cross-scenario generality experiments, we compare IGAA with service scheduling baseline methods designed for specific scenarios, including:
\begin{itemize}
    \item \textbf{IMAAC} \cite{ref14}: A service scheduling algorithm suitable for real-time edge computing, capable of allocating CPU, DPU, GPU, and other computing resources, trained under a multi-agent actor-critic framework.
    \item \textbf{PPOCO} \cite{ref15}: A method for edge VR service scheduling, modeling VR applications as directed acyclic graphs and jointly optimizing resource allocation, offloading strategies, and system costs.
    \item \textbf{TSS} \cite{ref16}: Designed for resource-constrained scenarios such as IoT and vehicular networks, optimizing service completion time and resource utilization based on a unique instantaneous reward mechanism.
    \item \textbf{Best Effort (BE)}: This method is a simple method with certain generality, which allocates resources to each service based on its minimum resource requirements and selects the nearest available edge node to execute the service.
\end{itemize}
\subsection{Generality Experiment}
\subsubsection{Cross-Scenario Generality}

\begin{table*}[htbp]
\centering
\caption{Cross-Scenario Generality of Different Methods}
\includegraphics[width=6in]{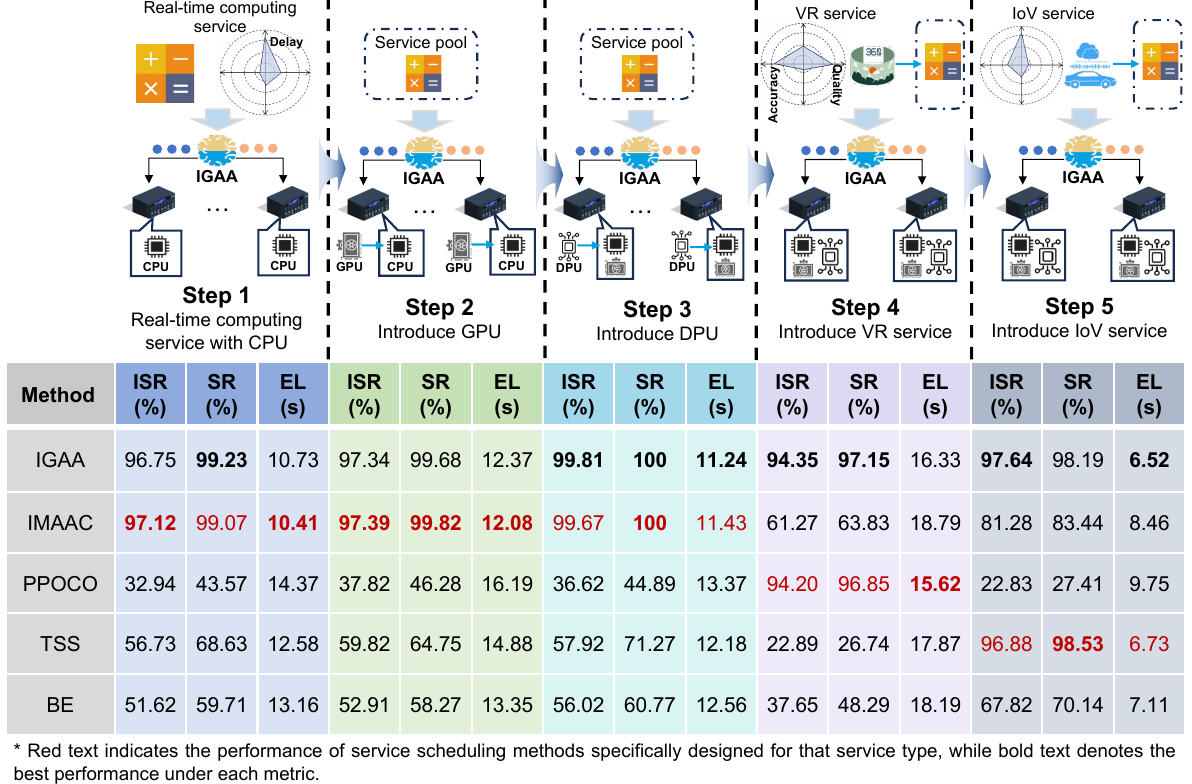}
\end{table*}

First, we examine IGAA's service scheduling performance across multiple scenarios and compare it with scenario-specific scheduling algorithms. As shown in Table II, during the training process, we initially set up a scenario with only CPU, storage resources, and real-time computing services. Then we gradually introduce GPU and DPU resources, VR services, and vehicular network services. There is high heterogeneity among these resources and the intents of the services, which better demonstrates the generalizability of IGAA. We evaluate the trained IGAA in the five scenarios above and compare it with scenario-specific service scheduling methods. We measure metrics such as user intent satisfaction rate (ISR), task success rate (SR), and average execution latency (EL). Considering that the scenario-specific service scheduling methods mentioned earlier cannot understand user intent, for fairness we integrate the same LLM to obtain user intent vectors for them. The comparison results are shown in Table II. IGAA’s performance metrics are close to those of the scenario-specific service scheduling methods across various scenarios. For some relatively less favorable results of IGAA, the maximum differences in the three metrics are 3.81\%, 3.45\%, and 4.55\%, which we consider acceptable. Moreover, it is evident that when these scenario-specific service scheduling methods are applied to other scenarios, they suffer from severe performance degradation.

\subsubsection{Adaptability to New Scenarios}
\newcolumntype{L}{>{\raggedright\arraybackslash}X}
\newcolumntype{C}{>{\centering\arraybackslash}X}

\begin{table}[htbp]
\centering
\caption{Adaptability to New Scenarios of Different Methods}
\label{tab:method_comparison_step6}
\begin{tabularx}{\columnwidth}{
  >{\hsize=1.5\hsize}L  
  >{\hsize=1.1\hsize}L   
  >{\hsize=0.8\hsize}C   
  >{\hsize=0.8\hsize}C   
  >{\hsize=0.8\hsize}C   
}
\specialrule{2pt}{0pt}{0pt}
\textbf{Step} & \textbf{Method} & \textbf{ISR (\%)} & \textbf{SR (\%)} & \textbf{EL (s)} \\
\midrule

\textit{Step 6: Introduce image detection service}
 & IGAA-step 6 & 97.31 & 98.72 & 15.37 \\
\midrule 
\textit{Introduce an image detection service in the real-time computing scenario}
 & IMAAC & 67.23 & 72.82 & 18.74 \\
\midrule 
\textit{Introduce an image detection service in the VR scenario}
 & PPOCO & 81.64 & 87.26 & 16.79 \\
\midrule 
\textit{Introduce an image detection service in the IoV scenario}
 & TSS & 32.87 & 37.71 & 16.61 \\
\midrule 
\textit{Introduce image detection service}
 & BE & 45.51 & 49.84 & 17.67 \\
\specialrule{2pt}{0pt}{0pt}
\end{tabularx}
\end{table}

We also examine the gap between IGAA and other methods in adapting to new scenarios to validate IGAA's generalizability. The introduction of new resource types in a scenario creates inevitable dimension mismatches. Consequently, other models cannot adapt without external modifications to their training code or structure. Therefore, when evaluating adaptability to new scenarios, we only consider the introduction of new service types. In Step 6, we introduce an image detection service into the Step 5 scenario and the aforementioned scenarios and retrained all models. In the testing environment, the new service is mixed with the existing services in each scenario, and the performance metrics of various methods are shown in Table III. All three scenario-specific methods experience significant performance degradation, which is because these methods lack sufficient generalizability. On the new dataset, the absence of proper reward function guidance prevents the models from accurately capturing the demands of the new service. Consequently, they fail to make correct service scheduling decisions. Furthermore, the original service scheduling strategies were also negatively affected. Among them, PPOCO shows the smallest performance drop, while TSS suffers the largest. This is because the scheduling logic of the image detection service is similar to that of the VR service, both requiring high graphics processing capability, i.e., substantial GPU resources, while being relatively insensitive to service latency. TSS, however, excels in vehicular network scenarios where latency sensitivity is high and is less capable of capturing the GPU resource requirements of services, hence the largest performance decline. IGAA, on the other hand, maintains high levels across all three metrics, demonstrating sufficient generalizability to adapt to new scenarios. Even compared with PPOCO, IGAA achieves a 19.19\% improvement in intent satisfaction rate.

\subsubsection{Scalability Test}
\begin{figure}[t]
\centering
\includegraphics[width= \columnwidth]{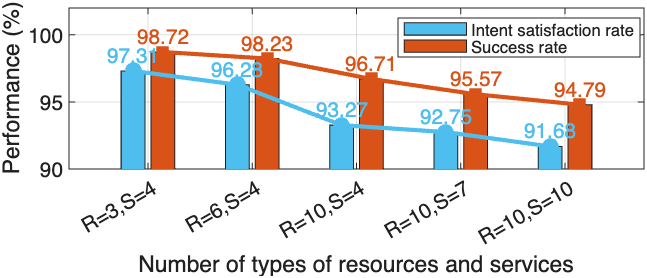}
\caption{Scalability analysis of the IGAA framework. Even when $R$ = 10 and $S$ = 10, both intent satisfaction rate and sucess rate remain above 90\%}
\end{figure}
To validate IGAA’s scalability, we progressively increase the number of resource types ($R$) and service types ($S$) and observe the corresponding changes in IGAA’s performance. Considering the large latency differences among different services, here we focus only on the more general metrics: intent satisfaction rate and task completion rate. As shown in Fig. 5, as the number of resources and service types increases, both intent satisfaction rate and task success rate gradually decline, with the increase in resource types causing a more pronounced performance drop. We believe this is because introducing too many resources and services makes the network overly complex, increasing the difficulty of convergence while limiting the ability to extract specific features. Since our proposed APOTL is designed to train a new sub-network, adding new services only affects the parameters of the shared network. In contrast, in RCETL, introducing new resources affects more parameters, leading to a more significant performance decline. Nevertheless, even when $R$ = 10 and $S$ = 10, both metrics remain above 90\%, which we consider sufficient for most edge scenarios. In fact, most edge nodes do not possess so many schedulable resource types. Our experiments consider some more specialized resource configurations or finer-grained resource types, such as dedicated I/O interfaces and secure hardware modules.
\subsection{Ablation Experiment}
In this section, we demonstrate the contributions of the various components designed within the IGAA framework through ablation experiments. 
\subsubsection{The Effect of IGAA Transfer Learning}
First, we examine the role of the IGAA transfer learning algorithm in improving model learning efficiency and performance. It should be noted that during IGAA training, multiple metrics are often used, especially task-specific metrics, to optimize model performance. Here, we focus only on representative metrics that are applicable to almost all services: training reward, task success rate, and latency, to illustrate the effect of the IGAA transfer learning algorithm. We repeat the training from Step 2 to Step 5 in Subsection VI. A and compared the performance of IGAA transfer learning against normal training to observe the differences.
\begin{figure*}[tpb]
\centering
\subfloat[]{\includegraphics[width=1.2in]{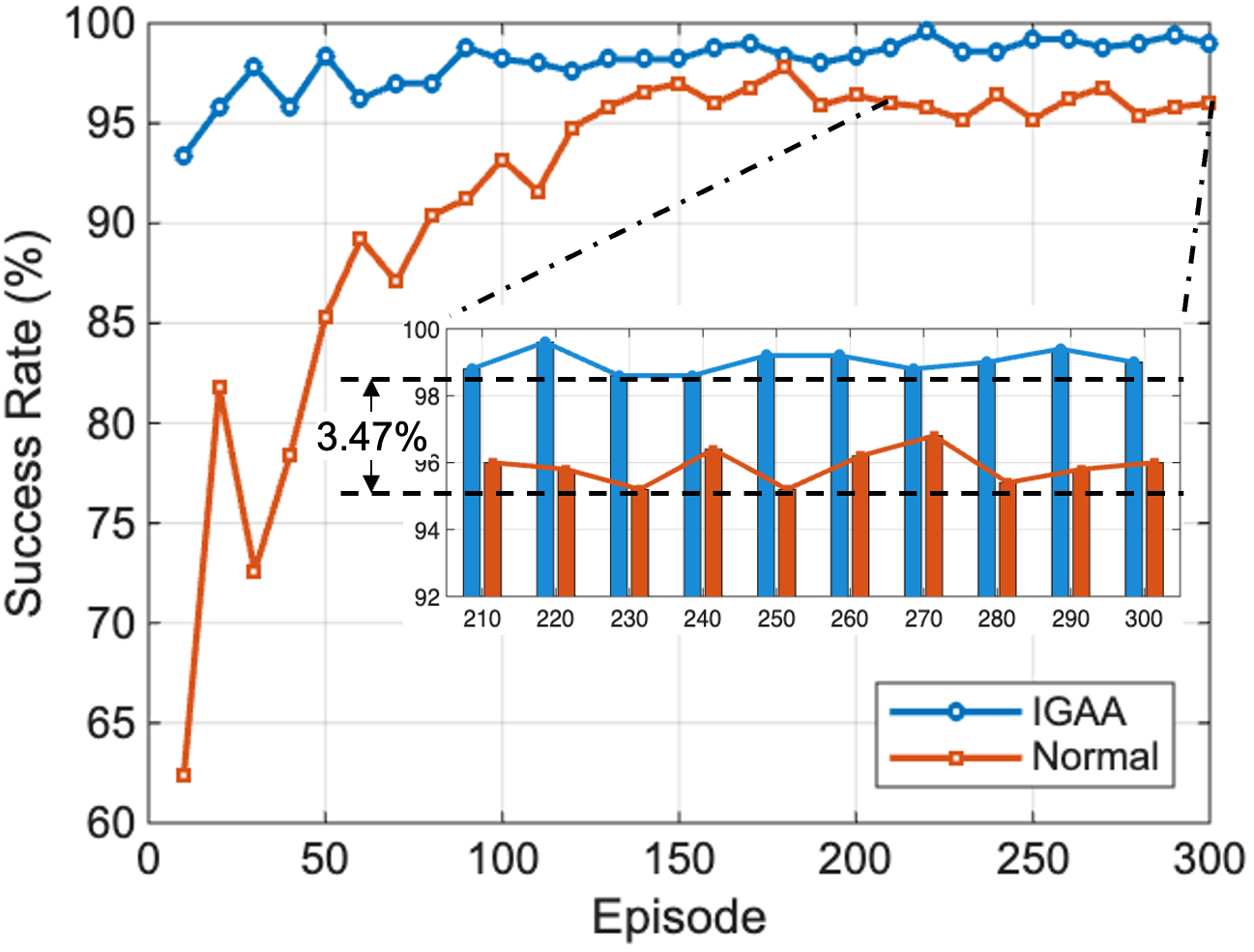}%
}
\subfloat[]{\includegraphics[width=1.2in]{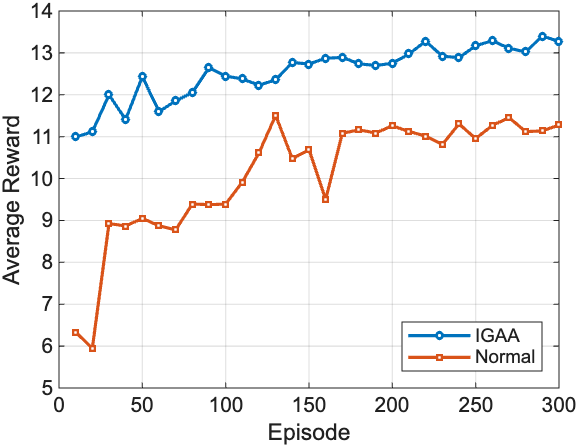}%
}
\subfloat[]{\includegraphics[width=1.2in]{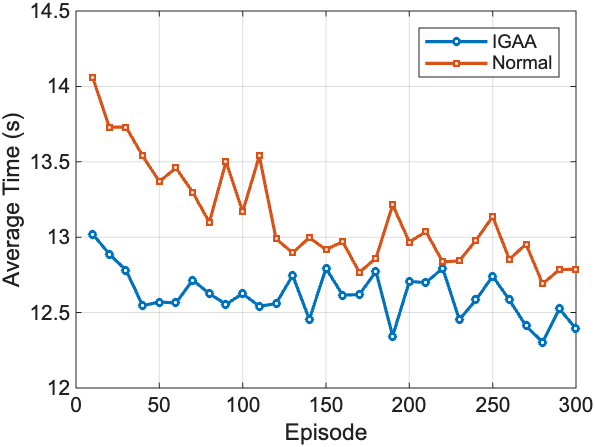}%
}
\subfloat[]{\includegraphics[width=1.2in]{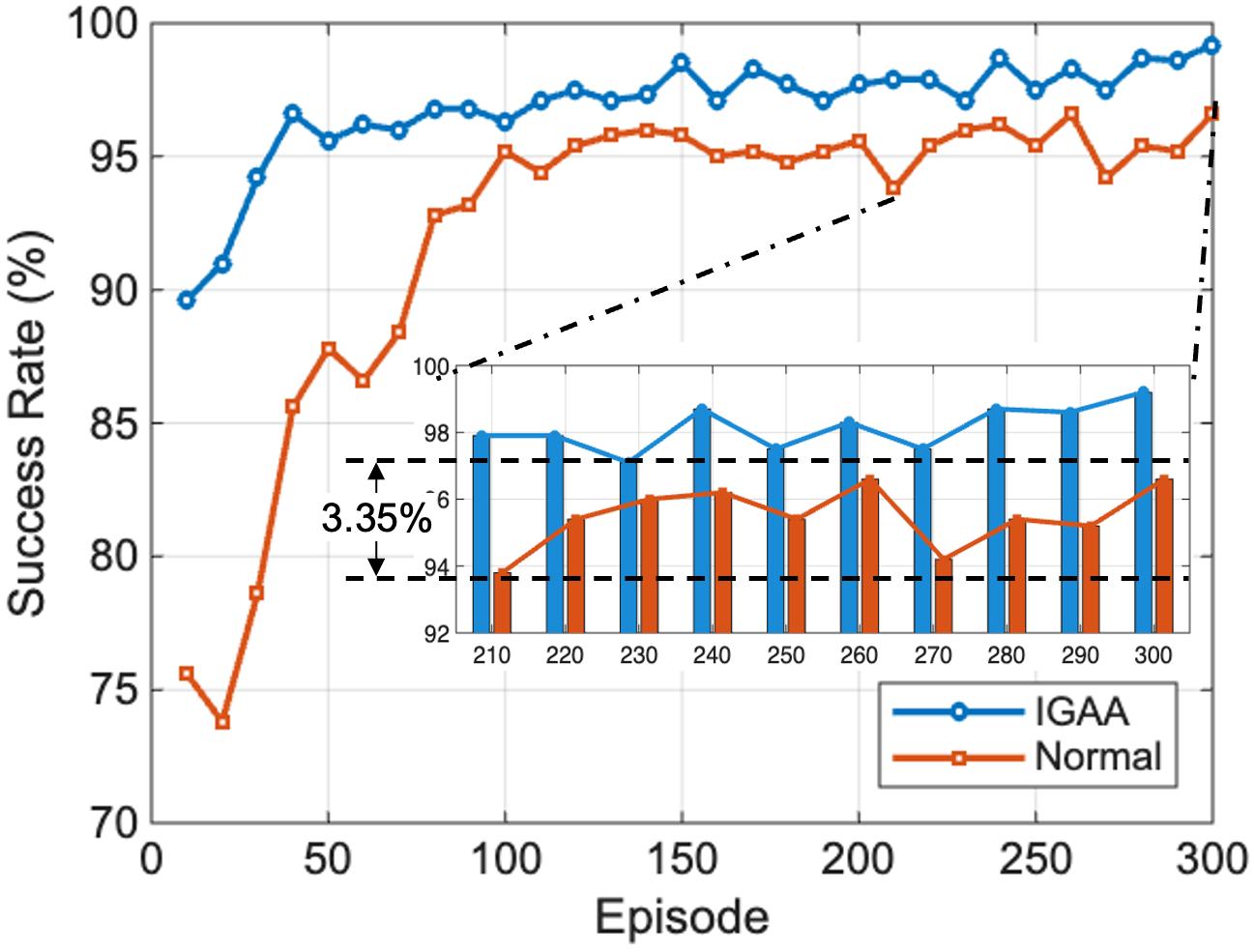}%
}
\subfloat[]{\includegraphics[width=1.2in]{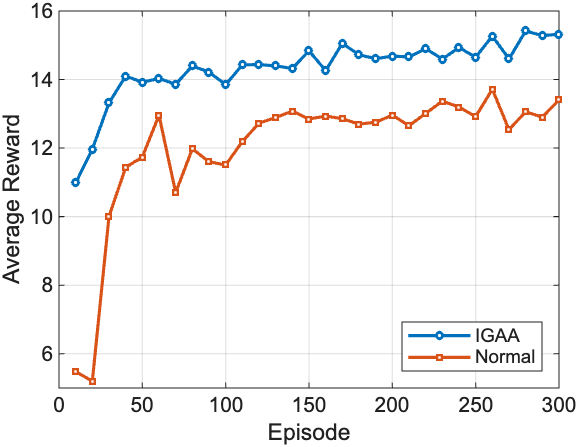}%
}
\subfloat[]{\includegraphics[width=1.2in]{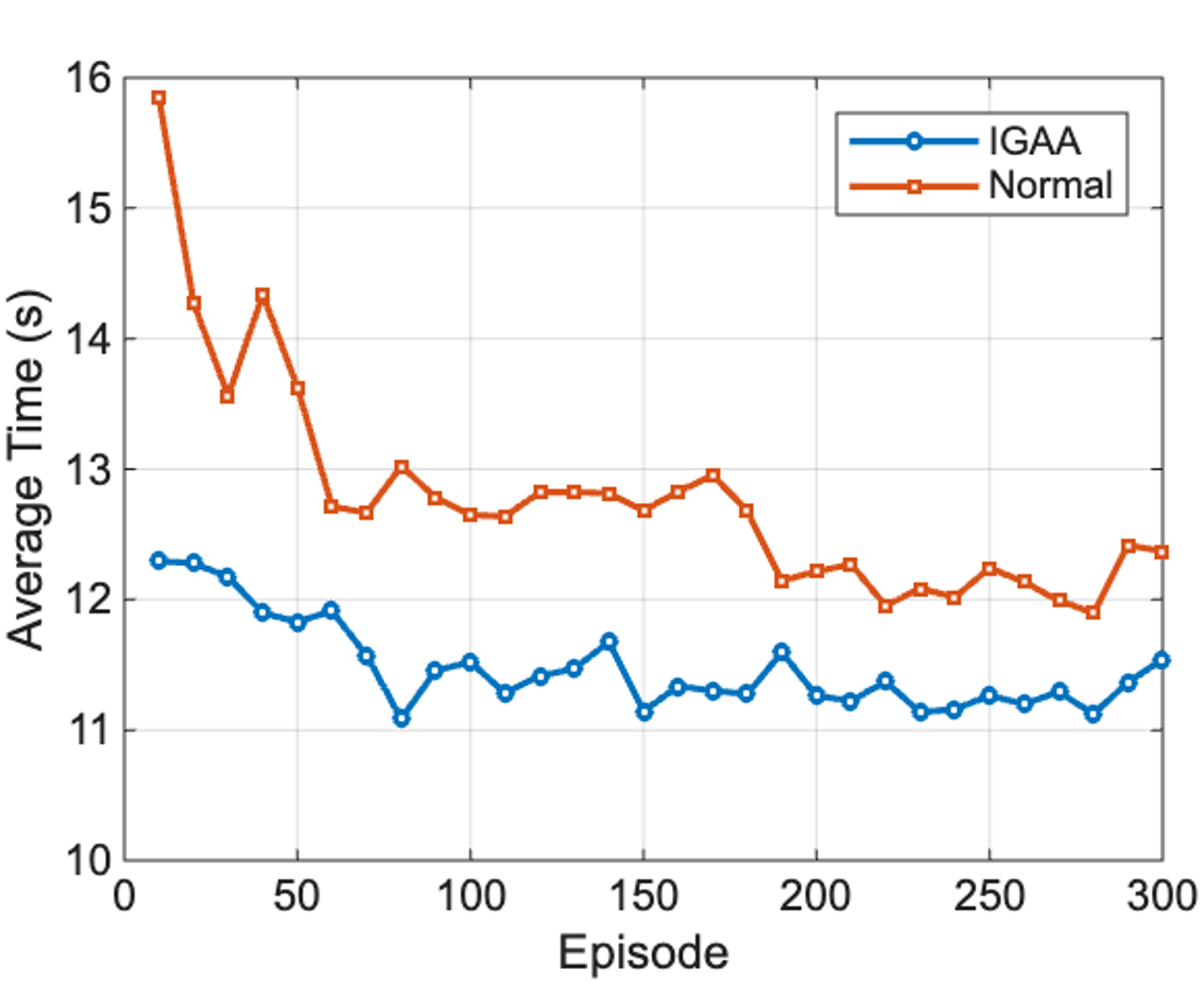}%
}
\\
\subfloat[]{\includegraphics[width=1.2in]{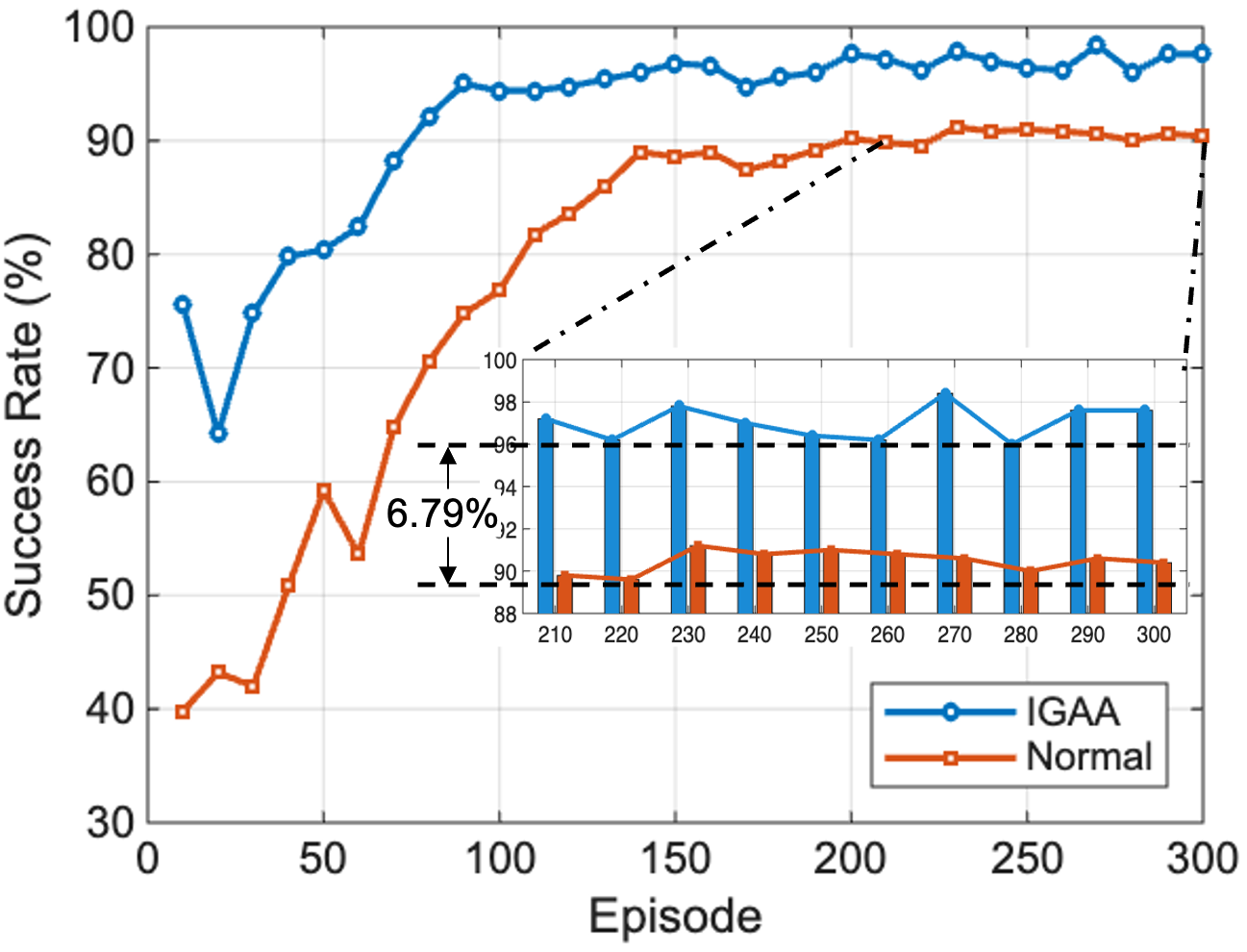}%
}
\subfloat[]{\includegraphics[width=1.2in]{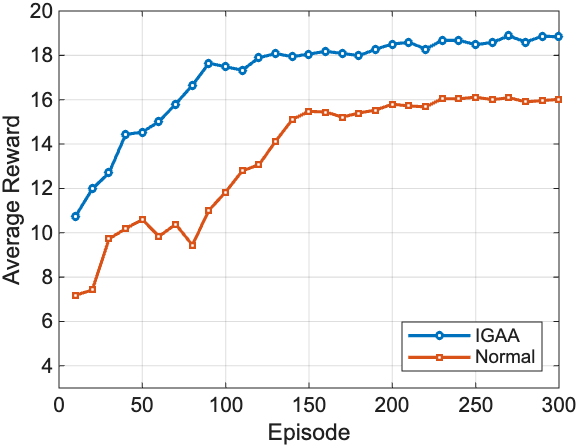}%
}
\subfloat[]{\includegraphics[width=1.2in]{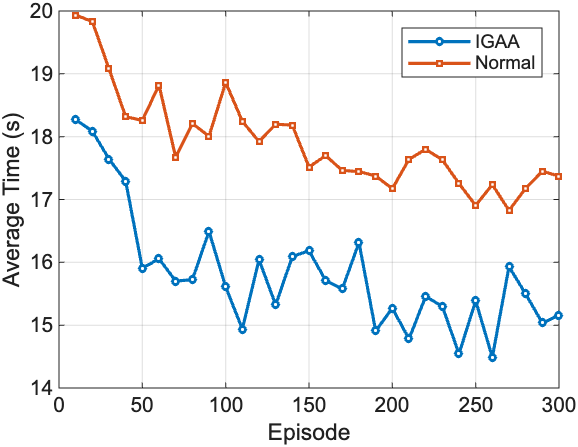}%
}
\subfloat[]{\includegraphics[width=1.2in]{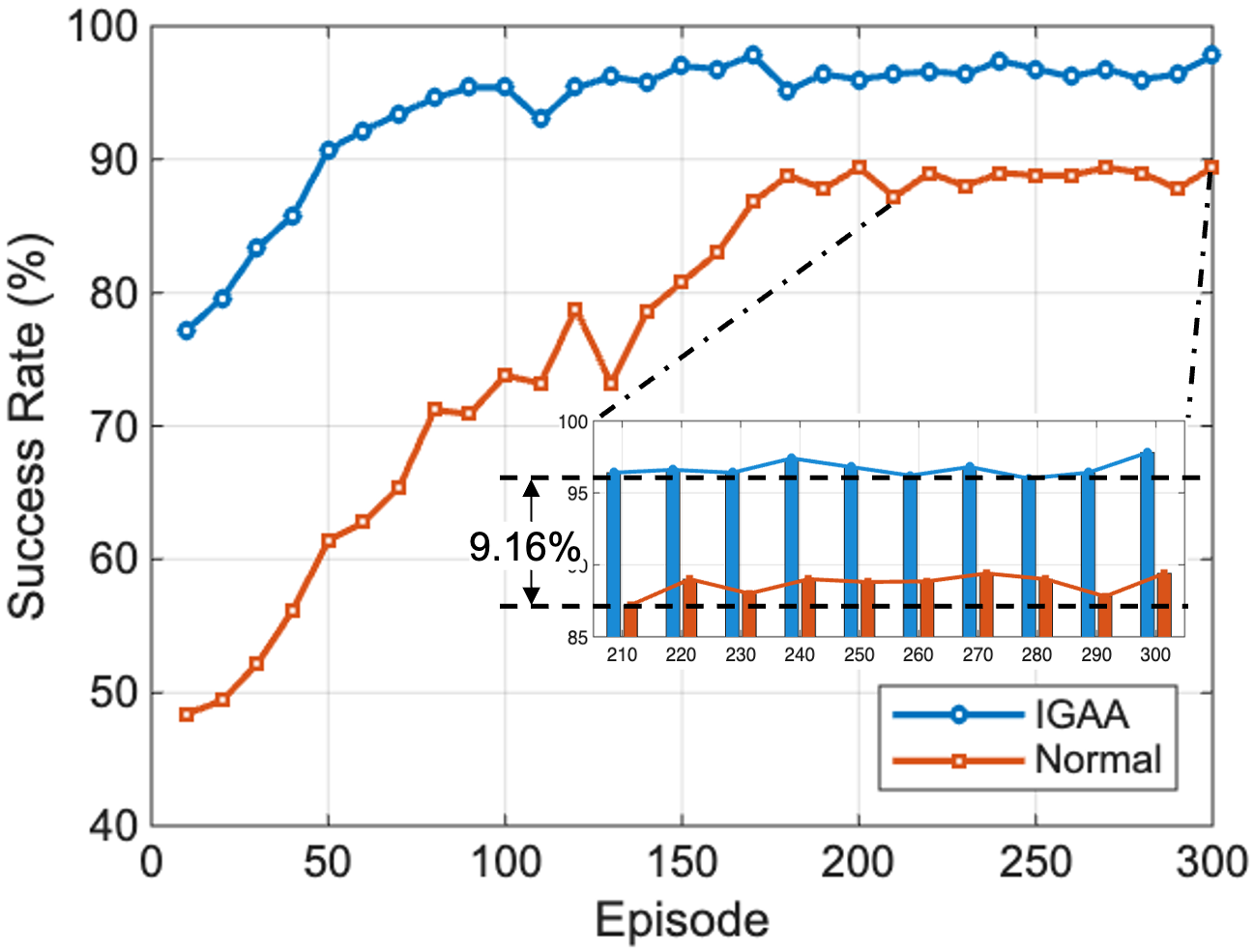}%
}
\subfloat[]{\includegraphics[width=1.2in]{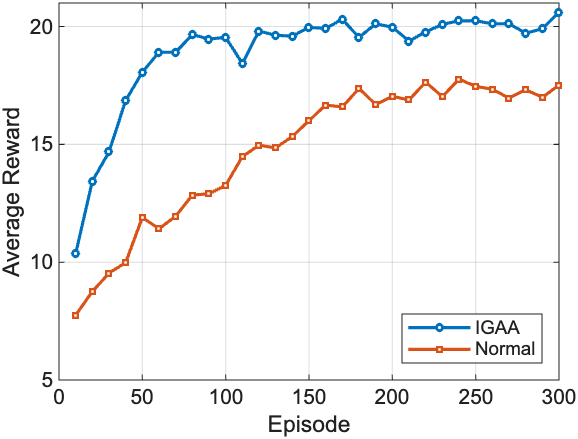}%
}
\subfloat[]{\includegraphics[width=1.2in]{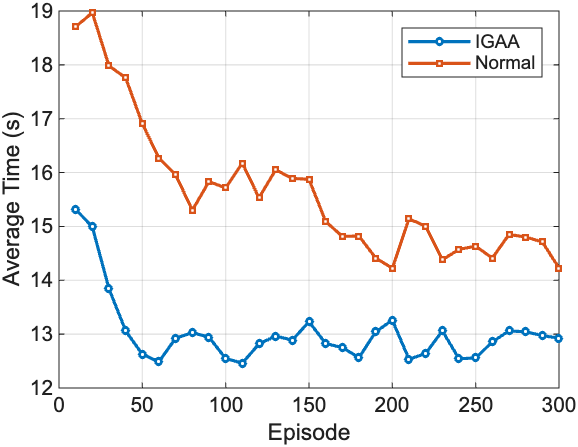}%
}
\caption{The differences between the IGAA transfer learning algorithm and normal training in terms of task success rate, training reward, and average task execution time are shown. Panels (a)–(c) illustrate the differences in Step 2, (d)–(f) show the differences in Step 3, (g)–(i) show the differences in Step 4, and (j)–(l) show the differences in Step 5.}
\end{figure*}

As shown in Fig. 6, IGAA's transfer learning algorithm consistently converges faster and achieves better performance compared to normal training methods. The gap becomes increasingly pronounced as the scenarios grow more complex. This is because normal training lacks reference from existing service scheduling experience, resulting in slower learning of effective strategies. Moreover, without analyzing relationships between new and existing scenarios, conventionally trained models exhibit weaker feature extraction capabilities for key aspects. This limitation leads to lower performance upon convergence compared to IGAA's transfer learning algorithm. To illustrate this point more concretely, we present two examples that demonstrate the advantages of the two sub-algorithms in capturing the relationships between new and existing scenarios.

\begin{figure}[h]
\centering
\includegraphics[width= 2.5in]{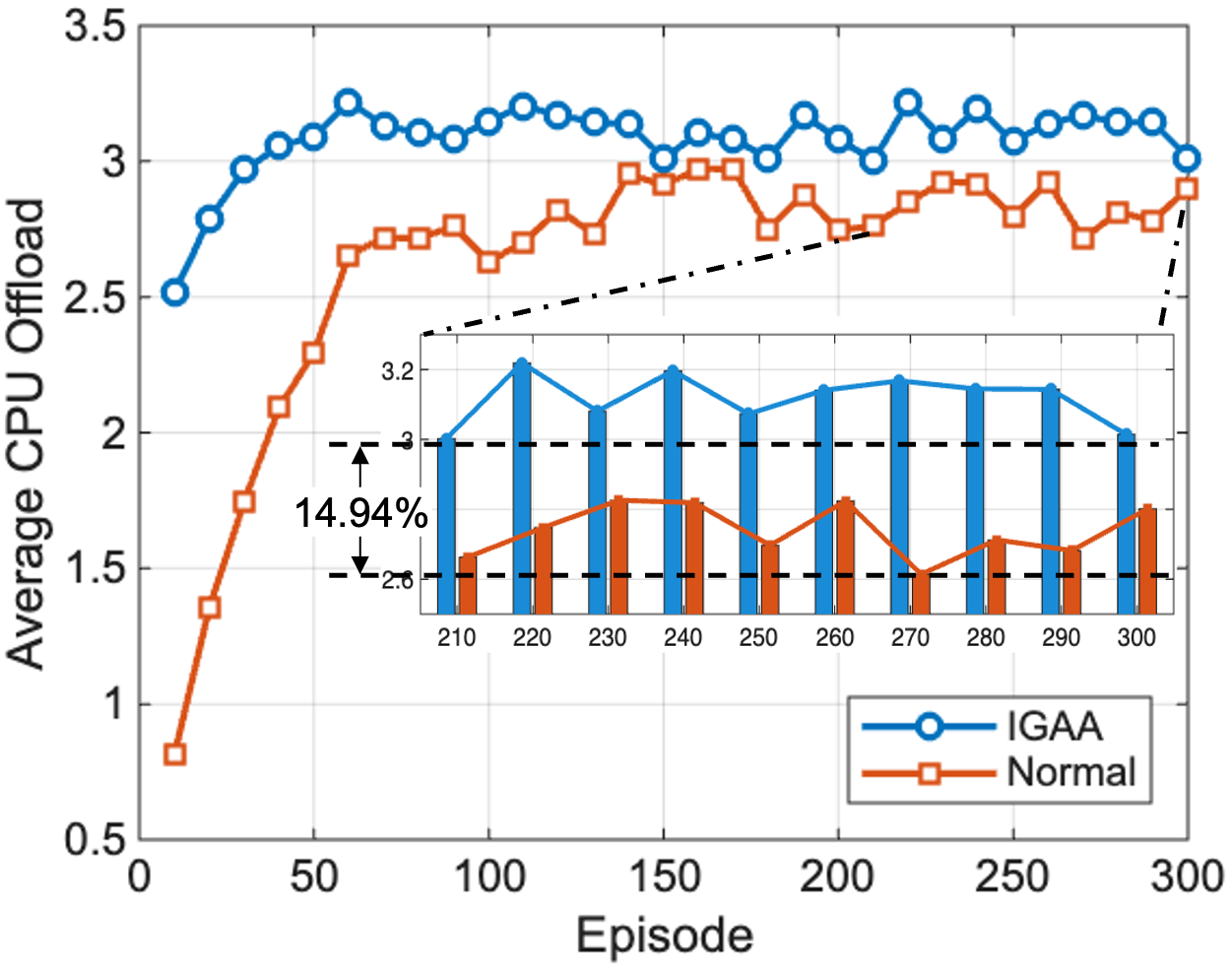}
\caption{The differences between the IGAA transfer learning algorithm and normal training in terms of average CPU offloading}
\end{figure}

We first examine the RCETL algorithm. In Step 3, we introduce the DPU, which can offload part of the CPU’s data processing tasks and reduce task execution latency. To evaluate the difference between RCETL and normal training, we observe the average offloading metric in Step 3. In this work, we define offloading as the amount of CPU resources saved by offloading tasks to the DPU. As shown in Fig. 7, RCETL achieves a high offloading level from the beginning and consistently maintains an advantage on this metric. This is because RCETL captures the cooperative features between CPU and DPU by analyzing causal relationships among resources. This allows the algorithm to focus on optimizing parameters related to them and thereby learning a more effective CPU-DPU collaborative scheduling strategy. Next, we examine the role of the APOTL algorithm. We observe that in Fig. 6(g), IGAA’s convergence speed is slower than in Fig. 6(j). This is because the vehicular network service introduced in Step 5 is similar to the real-time computing services in existing scenarios, both requiring high latency sensitivity. APOTL can accurately capture this and leverage the scheduling strategies of real-time computing services during training. In contrast, the VR service introduced in Step 4 differs substantially from real-time computing services. APOTL can only draw on some basic resource allocation strategies, resulting in slower convergence compared to Step 5.

\subsubsection{The Effect of GIR}
To validate the role of GIR, we repeat the steps from Fig. 5 and set up a version of IGAA without the GIR process as a baseline. We then observe the changes in intent satisfaction rate for real-time computing services and VR services at each step of the test.

\begin{figure}[h]
\centering
\includegraphics[width= \columnwidth]{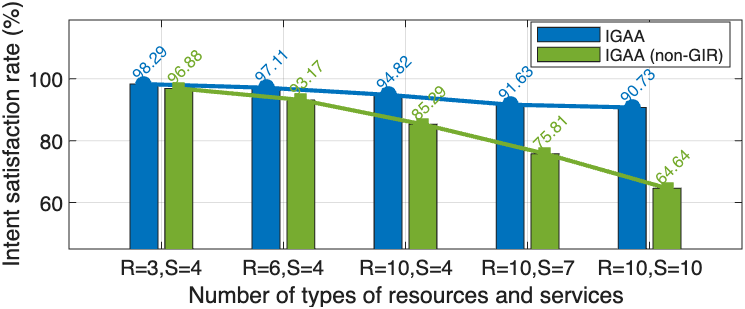}
\caption{The effect of GIR in real-time computing services scheduling}
\end{figure}
\begin{figure}[h]
\centering
\includegraphics[width= \columnwidth]{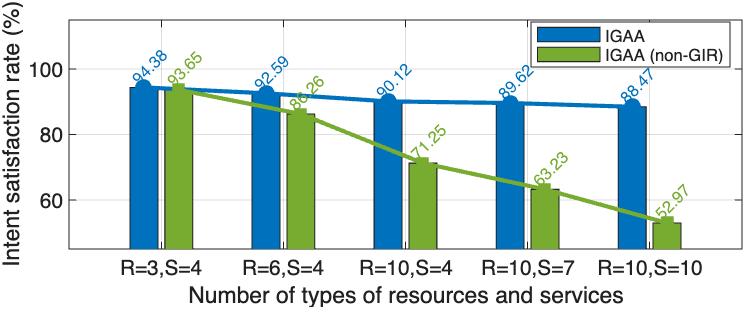}
\caption{The effect of GIR in VR services scheduling}
\end{figure}

As shown in Figs. 8 and 9, the effect of GIR is not obvious during the initial stage, and the model can still maintain a high intent satisfaction rate. This is because, at this stage, the number of training iterations and the input dimensions are relatively small, so learning new scheduling strategies does not significantly impact existing strategies. As the number of resources and service types increases, new scheduling strategies begin to affect existing ones. Without training with GIR, the intent satisfaction rates for both services start to decline. The VR service is more affected because the CPU and GPU resources and scheduling strategies involved in real-time computing services are relatively general. In subsequent learning, these general strategies still form the foundation, so only some specific resource or service scheduling logic affects real-time computing services. In contrast, VR services rely on specific strategies that require high graphics processing. When the model begins to learn scheduling strategies for services with other requirements, the parameters of the model without GIR drift toward the objectives of the new tasks. IGAA learns the scheduling strategies of new services through a new subtask layer. However, the shared parameter layer is still affected, leading to performance degradation on VR services, i.e., catastrophic forgetting. In contrast, IGAA with GIR continually learns from GIR data, ensuring that while adapting to new scenarios, the scheduling capabilities for existing services remain at a high level.

\subsubsection{The Effect of Evaluation and Correction Model}
To validate the effect of the evaluation and correction model, we pause IGAA's training after the LLM generated the training code for Step 2. We then manually modified the reward function in the training code to induce IGAA to overuse the CPU. This simulated a scenario where the LLM hallucinated and set an incorrect service scheduling logic. We train versions of IGAA with and without the evaluation and correction model and observe their performance in the test environment.

\begin{figure}[t]
\centering
\includegraphics[width= \columnwidth]{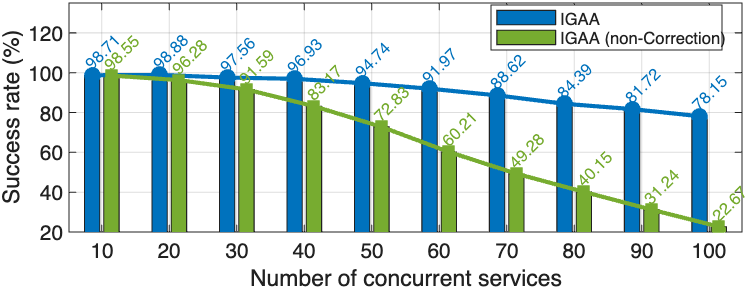}
\caption{The effect of the evaluation and correction model}
\end{figure}

As shown in Fig. 10, we gradually increase the number of concurrent tasks in the test environment. In the initial stage, the service success rates of both versions were similar. However, as the number of tasks increases, the service success rate of IGAA without the evaluation and correction model began to decline. This is because it incorrectly allocates excessive CPU resources; with few concurrent services, the impact was limited. As concurrency increases, the CPU became a bottleneck, resulting in lower service success rates. In contrast, IGAA, with the evaluation and correction model, can detect this error during training. It utilizes metrics such as intent satisfaction rate and the utilization of similar resources for detection. Upon detection, it promptly adjusts the reward function, thereby producing a service scheduling model with better performance.

\section{Conclusion}
In this work, we have proposed a novel intent-driven General Edge Service Scheduling AAI framework. In this framework, we have leveraged a generative meta-learning paradigm to enable the AAI to autonomously engage in learning from simple scenarios continually, ultimately achieving generalized edge service orchestration. We have integrated IGAA with three core mechanisms. First, we introduce an N-S-I matrix mapping method to transform complex service relationships into computable numerical associations. This format assists LLMs in simulating new scenarios and generating datasets. 
Second, we have developed an easy-to-hard generalization learning framework. We have integrated combined with customized RCETL and APOTL algorithms, enabling the service scheduling model to effectively leverage prior knowledge to adapt to new environments. Finally, to mitigate LLM hallucinations and prevent catastrophic forgetting, we have integrated a scenario evaluation and correction model and designed the GIR mechanism. Extensive experimental results have demonstrated that the proposed IGAA framework achieves strong generalization and scalability across diverse edge network scenarios.

\bibliography{reference}
\bibliographystyle{IEEEtran}

\end{document}